\documentclass[aps,twocolumn,superscriptaddress,floatfix,nofootinbib,longbibliography, reprint]{revtex4-1}
\usepackage{graphicx,amsmath,amsfonts,amssymb,amsthm,xr}
\usepackage{epsfig,amsmath,amssymb,color,dsfont,upgreek,physics, bm}
\usepackage{mathrsfs}
\usepackage{mathtools}
\usepackage{bbold}
\usepackage{comment}
\usepackage{float}
\usepackage[bookmarks=true,colorlinks,linkcolor=Blue,urlcolor=Blue,citecolor=Blue]{hyperref}
\usepackage[dvipsnames]{xcolor}
\usepackage{orcidlink}

\renewcommand{\H}{\hat{\mathcal{H}}}
\newcommand{\hc}{\text{H.c.}}
\newcommand{\Zt}{$\mathbb{Z}_2$ }

\newcommand{\ad}{\hat{a}^{\dagger}}
\renewcommand{\a}{\hat{a}^{\phantom{\dagger}}}

\newcommand{\tauZ}{\hat{\tau}^{z}}
\newcommand{\tauX}{\hat{\tau}^{x}}

\newcommand{\veci}{\textbf{i}}
\newcommand{\vecj}{\textbf{j}}

\newcommand{\lef}{\left(}
\newcommand{\rgh}{\right)}
\newcommand{\ijB}{\langle \textbf{i} , \textbf{j} \rangle}

\begin{document}
\title{Matter-induced plaquette terms in a \texorpdfstring{$\mathbb{Z}_2$}{Z2} lattice gauge theory}

\author{Matja\v{z} Kebri\v{c}${}^{\ddagger}$\orcidlink{0000-0003-2524-2834}}
\email{matjaz.kebric@colorado.edu}
\affiliation{JILA and Department of Physics, University of Colorado, Boulder, Colorado 80309, USA}

\author{Fabian D\"oschl${}^{\ddagger}$\orcidlink{0009-0005-5067-004X}}
\affiliation{Department of Physics and Arnold Sommerfeld Center for Theoretical Physics (ASC), Ludwig-Maximilians-Universit\"at M\"unchen, Theresienstra\ss e 37, D-80333 M\"unchen, Germany}
\affiliation{Munich Center for Quantum Science and Technology (MCQST), Schellingstra\ss e 4, D-80799 M\"unchen, Germany}

\author{Umberto Borla\orcidlink{0000-0002-4224-5335}}
\affiliation{Department of Physics and Arnold Sommerfeld Center for Theoretical Physics (ASC), Ludwig-Maximilians-Universit\"at M\"unchen, Theresienstra\ss e 37, D-80333 M\"unchen, Germany}
\affiliation{Munich Center for Quantum Science and Technology (MCQST), Schellingstra\ss e 4, D-80799 M\"unchen, Germany}

\author{Jad C.~Halimeh\orcidlink{0000-0002-0659-7990}}
\affiliation{Department of Physics and Arnold Sommerfeld Center for Theoretical Physics (ASC), Ludwig-Maximilians-Universit\"at M\"unchen, Theresienstra\ss e 37, D-80333 M\"unchen, Germany}
\affiliation{Max Planck Institute of Quantum Optics, 85748 Garching, Germany}
\affiliation{Munich Center for Quantum Science and Technology (MCQST), Schellingstra\ss e 4, D-80799 M\"unchen, Germany}
\affiliation{Department of Physics, College of Science, Kyung Hee University, Seoul 02447, Republic of Korea}

\author{Ulrich Schollw\"ock\orcidlink{0000-0002-2538-1802}}
\affiliation{Department of Physics and Arnold Sommerfeld Center for Theoretical Physics (ASC), Ludwig-Maximilians-Universit\"at M\"unchen, Theresienstra\ss e 37, D-80333 M\"unchen, Germany}
\affiliation{Munich Center for Quantum Science and Technology (MCQST), Schellingstra\ss e 4, D-80799 M\"unchen, Germany}

\author{Annabelle Bohrdt\orcidlink{0000-0002-3339-5200}}
\affiliation{Department of Physics and Arnold Sommerfeld Center for Theoretical Physics (ASC), Ludwig-Maximilians-Universit\"at M\"unchen, Theresienstra\ss e 37, D-80333 M\"unchen, Germany}
\affiliation{Munich Center for Quantum Science and Technology (MCQST), Schellingstra\ss e 4, D-80799 M\"unchen, Germany}

\author{Fabian Grusdt\orcidlink{0000-0003-3531-8089}}
\email{fabian.grusdt@lmu.de}
\affiliation{Department of Physics and Arnold Sommerfeld Center for Theoretical Physics (ASC), Ludwig-Maximilians-Universit\"at M\"unchen, Theresienstra\ss e 37, D-80333 M\"unchen, Germany}
\affiliation{Munich Center for Quantum Science and Technology (MCQST), Schellingstra\ss e 4, D-80799 M\"unchen, Germany}

\begin{abstract} 
Lattice gauge theories (LGTs) provide a powerful framework for studying confinement, topological order, and exotic quantum matter.
In particular, the paradigmatic phenomenon of confinement, where dynamical matter is coupled to gauge fields and forms bound states, remains an open problem.
In addition, LGTs can provide low-energy descriptions of quantum spin liquids, which is the focus of ongoing experimental research.
However, the study of LGTs is often limited theoretically by their numerical complexity and experimentally in implementing challenging multi-body interactions, such as the plaquette terms crucial for the realization of many exotic phases of matter.
Here we investigate a $(2+1)$D $\mathbb{Z}_2$ LGT coupled to hard-core bosonic matter featuring a global U(1) symmetry, and show that dynamical matter naturally induces sizable plaquette interactions even in the absence of explicit plaquette terms in the Hamiltonian.
Using a combination of density matrix renormalization group simulations and neural quantum state calculations up to a system size of $20 \times 20$, we analyze the model across different fillings and electric field strengths.
At small coupling strength, we find a large plaquette expectation value, independent of system size, for a wide range of fillings, which decreases in the presence of stronger electric fields. Furthermore, we observe signatures of a confinement-deconfinement transition at weak coupling strengths.
Our results demonstrate that dynamical U(1) matter can induce complex multi-body interactions, suggesting a natural route to the realization of strong plaquette terms and paving the way for realizing a topological quantum spin liquid protected by a large gap.
\end{abstract}

\date{\today}
\maketitle

\begingroup
\renewcommand{\thefootnote}{\fnsymbol{footnote}}
\footnotetext[3]{These authors contributed equally to this work.}
\endgroup

\textit{Introduction.---}
The development of lattice gauge theories (LGTs) enabled an effective study of many-body systems in strongly coupled regimes, where one of the most prominent applications is the study of quark confinement \cite{Wilson1974, Kogut1979, Fradkin1979}.
In addition, LGTs emerge as effective low energy theories that describe exotic phases of matter such as topological spin liquids \cite{Wen2007, Sachdev2018, Semeghini2021}, fractionalization of charges \cite{Senthil2000, Sedgewick2002, Gazit2017}, and high-$T_c$ superconductivity \cite{Demler2002, Kaul2007}.
Furthermore, the toric code, which is an LGT in one of its simplest forms, has been proposed  as a basis for fault tolerant quantum computing \cite{Kitaev2003}.
Despite the tremendous progress in the study of LGTs \cite{Fradkin1979}, especially using quantum Monte Carlo based methods for systems that contain static charges \cite{CarmenBanuls2020}, the confinement problem of dynamical matter at finite filling is still not fully understood \cite{Greensite2011}.
This is also in part due to the complex nature of LGTs, which often makes large-scale numerical treatment of LGTs difficult beyond the simplest one-dimensional cases.
There, one can employ the density-matrix renormalization group (DMRG) method to find the ground state of the system \cite{Schollwoeck2011, White1992}.
Although DMRG can be extended to quasi two-dimensional systems on a cylinder with limited circumference \cite{Schollwoeck2011}, quantum Monte Carlo based methods are the preferred choice in large-scale two-dimensional calculations \cite{CarmenBanuls2020}.
However, for arbitrary matter filling the sign problem makes the calculations especially challenging \cite{Troyer2005, Gazit2017, Assaad2016, CarmenBanuls2020}.
Furthermore, quantum Monte Carlo based methods become difficult to apply for real-time dynamics and one has to employ tensor network---e.g. matrix-product state (MPS)---based methods, whose power is limited in dimensions $D\geq2$ \cite{Banuls2020, Magnifico2025}.

A novel numerical approach that can overcome certain limitations of quantum Monte Carlo and tensor networks are neural quantum states (NQS) \cite{Carleo2017}.
They have demonstrated remarkable efficiency in representing a wide range of quantum states \cite{Carleo2017, Chen2024, Pfau2024, Szabo_2020, Chen_20202_sign, Lange_2024_RNN, Doeschl_2025, Kufel_2025, Moreno_2022, Viteritti2023,Roth2025, Lange_2024, Denis2025accurateneural, moss2025} and their dynamical properties \cite{Schmitt_2020, Guti_rrez_2022, Donatella_2023, Walle_2024, sinibaldi_2025, Schmitt_2025}.
Hence, NQS constitute an empirically successful alternative for the numerical simulation of two-dimensional quantum systems, with the potential to represent highly entangled states \cite{Gao_2017,Deng_Quantum_Ent_2017,Sharir_2022,Passetti2023}.

The development of quantum simulators has opened up a complementary approach to studying LGTs \cite{Halimeh2025, Banuls2020, Zohar2015, Wiese2013, Bauer2023, Bauer2023qSimHEP, DiMeglio2024, Halimeh2025QsimOutEquil}.
Quantum simulators based on trapping cold atoms in optical lattices or tweezers are highly tunable and can be used to investigate various quantum many-body models across different parameter regimes and lattice geometries.
With single-atom resolution, they can also provide snapshots of the system, and one can study ground state as well as dynamical properties of the models.
First building blocks and one-dimensional LGTs have already been realized \cite{Barbiero2019, Goerg2019, Schweizer2019, Mil2020, Yang2020, Zhou2022}, including the extension beyond one dimension to a mixed-dimensional setting \cite{Kebric2025}.
Achieving a fully two-dimensional large-scale analog quantum simulation of LGTs, which is one of the ultimate goals, is thus an active area of ongoing research.
At the heart of the problem lies the difficult implementation of strong multi-body interactions, such as the plaquette terms, which can be introduced as weak multi-order perturbative terms \cite{Homeier2023, Paredes2008, Dai2017, Tian2025}.
These terms are important because they drive the deconfinement transition \cite{Fradkin1979}, and can stabilize quantum spin liquids \cite{Savary2016}.
In addition to analog simulators, emergent digital quantum computing platforms, which mainly rely on Trotterized time evolutions, recently enabled the study of confinement and dynamics in LGTs as hardware becomes increasingly more reliable \cite{Martinez2016, Gyawali2024, Cochran2025, Mildenberger2025, GonzalezCuadra2025, De2024}.

Here we study the emergence of magnetic plaquette terms in a two-dimensional ($2+1$D) \Zt LGT coupled to dynamical U(1) matter using DMRG \cite{Schollwoeck2011, White1992, hubig:_syten_toolk, hubig17:_symmet_protec_tensor_network}, and NQS \cite{Carleo2017, Passetti2023, Chen2024, Pfau2024}.
We first compute the ground state of the system using well-established DMRG simulations.
We complement these results by employing NQS to investigate larger system sizes, using DMRG as a benchmark.
Our results reveal a large expectation value of the \Zt plaquette operators, despite the absence of bare \Zt plaquette terms in the Hamiltonian, which is driven by the finite U(1) matter coupled to the gauge field.
This indicates that finite matter filling is sufficient to generate states involving finite plaquette terms, thus eliminating the need for their explicit implementation in quantum simulators.
This is in particular important in the realization of elusive quantum spin liquids in systems without native plaquette interactions \cite{Semeghini2021, Wang2025}.
Without matter, such systems exhibit small magnetic energy scales, below practically achievable temperature scales, and become classical spin liquids for temperatures below the larger energy scale of electric excitations \cite{Wang2025}.
As we show here, dynamical U(1) matter increases the magnetic energy scale, allowing the stabilization of quantum spin liquids.

\begin{figure}[t]
\centering
\epsfig{file=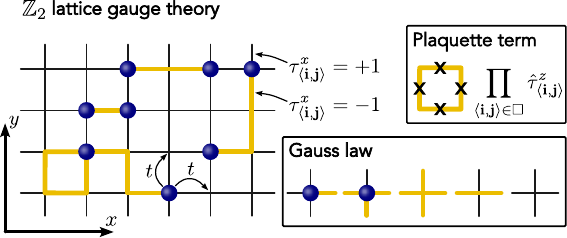, width=0.48\textwidth}
\caption{$(2+1)$D \Zt Lattice gauge theory coupled to matter. Hard-core bosons (blue spheres) defined on a two-dimensional (2D) lattice are coupled to a \Zt gauge field defined on the lattice links. 
We label the configuration of the \Zt electric field, also defined on the links, by drawings a yellow line on the link for $\tau^{x}_{\langle \veci, \vecj \rangle} = -1$, and leaving the link blank for $\tau^x_{\langle \veci, \vecj \rangle} = +1$.
The magnetic plaquette term is defined as a product of the gauge fields $\tauZ_{\langle \veci, \vecj \rangle}$ on the links of a single plaquette.
We consider the Gauss's law, Eq.~\eqref{eqGaussLaw}, in the physical sector which limits the possible configurations of the matter and the \Zt electric fields.
}
\label{FigOne}
\end{figure}

\textit{Model.---}
We consider a $(2+1)$D \Zt LGT coupled to U(1) matter
\begin{multline}
    \H = -t \sum_{\ijB} \lef \ad_{\veci} \tauZ_{\ijB} \a_{\vecj} + \hc \rgh \\ + \mu \sum_{\vecj} \lef \hat{n}_{\vecj} - \frac{1}{2} \rgh -h \sum_{\ijB} \tauX_{\ijB},
    \label{eq_2DLGTDef}
\end{multline}
where $\ad_{\vecj}$ $(\a_{\vecj})$ are hard-core boson creation (annihilation) operators defined on the sites of the square two-dimensional lattice $\vecj = (j_x, j_y)$; see Fig.~\ref{FigOne}.
We define the boson (parton) number operator as $\hat{n}_{\vecj} = \ad_{\vecj} \a_{\vecj}$, and include the chemical potential term $\propto \mu \sum_{\vecj} \hat{n}_{\vecj}$, to control the filling of the system in the DMRG calculations.
The \Zt gauge field $\tauZ_{\ijB}$ is defined on the links between neighboring lattice sites, $ \veci - \vecj = \pm a \textbf{e}_\mu$, where $a$ is the lattice spacing and $\textbf{e}_\mu$ is the unit vector along the two spatial directions $\mu = x,y$.
Partons are coupled to the gauge field via the nearest-neighbor hopping term, $\propto t$, and the last term in Eq.~\eqref{eq_2DLGTDef} is the \Zt electric field term, $\propto h$.
The $(2+1)$D \Zt LGT in Eq.~\eqref{eq_2DLGTDef} generally also contains the bare plaquette terms, which are defined as
\begin{equation}
    \H_{\square} = - J \sum_{\square} \prod_{\ijB \in \square} \tauZ_{\ijB},
    \label{eq_Z2Plaquette}
\end{equation}
where the product contains the \Zt gauge fields that form a single plaquette; see Fig.~\ref{FigOne}.
In this work, we demonstrate that finite plaquette term values are induced by the U(1) matter in the absence of the bare term in Eq.~\eqref{eq_Z2Plaquette}, i.e., when $J=0$.

The Gauss's law is defined by a set of local operators \cite{Prosko2017}
\begin{equation}
    \hat{G}_{\vecj} = \lef -1 \rgh^{\hat{n}_{\vecj}} \prod_{\textbf{k}: \left \langle \textbf{j}, \textbf{k} \right \rangle} \tauX_{\left \langle \textbf{j}, \textbf{k} \right \rangle},
    \label{eqGaussLaw}
\end{equation}
which commute with the Hamiltonian $\left [ \H, \hat{G}_{\vecj} \right ] = 0$ \cite{Prosko2017}.
The product of the \Zt electric field operators runs over all links connected to the site $\vecj$.
The Gauss's law $\hat{G}_{\vecj} \ket{\psi} = g_{\vecj} \ket{\psi}$ divides the Hilbert space into different sectors defined by the set of eigenvalues $\{ g_{\vecj}\}$.
We consider the physical sector, $\hat{G}_{\vecj} \ket{\psi} = + \ket{\psi}$, $\forall \vecj$, which corresponds to the Hilbert sector without background \Zt charges.
This allows us to integrate out the matter degrees of freedom, which simplifies our numerical calculations; see Supplemental Material~\cite{SM} for details.

%%%%%%%%%%%%%%%%%%%%%%%%%%%%%%%%%%%%%%%%%%%%%%%%%%%%%
%%%%%%%%%%%%%%%%%%%%%%%%%%%%%%%%%%%%%%%%%%%%%%%%%%%%%

We can define \Zt electric strings as links with a negative orientation of the \Zt electric field, $\tauX = -1$ \cite{Borla2020}.
Gauss's law Eq.~\eqref{eqGaussLaw}, provides an intuitive physical picture where matter is the source of the \Zt electric field; see Fig.~\ref{FigOne}.
More precisely, in the physical sector $g_{\vecj} = +1, \forall \vecj$, which we consider in this work, particles are attached to the ends of \Zt strings.
Although open strings can only start and end with matter, closed loops are also possible; see Fig.~\ref{FigOne}.
Using the string picture, it becomes evident that a finite value of the \Zt electric field term induces a linear confining potential in the strings $\propto 2 h \ell $, where $\ell$ is the length of the string \cite{Borla2020, Kebric2021}.
Hence, in the limit where $h \gg t, J$, the partons connected with a \Zt string confine into bound dimers (mesons).
We note that when $h = 0$, the model reduces to a system of free hard-core bosons, where the gauge fields are static, and the ground state energy is determined by the corresponding \Zt flux configuration.
In the case when $h = 0$, the hard-core bosons favor the zero-flux state, and are described by a quantum XY model \cite{Ding1992}.

The $(2+1)$D \Zt LGT model Eq.~\eqref{eq_2DLGTDef} conserves the total U(1) charge $\hat{N}^{a} = \sum_{\vecj} \hat{n}_{\vecj}$.
Thus, it differs from the model studied by Fradkin and Shenker \cite{Fradkin1979}, which includes pair creation and annihilation terms ${\propto \lef \ad_{\veci} \tauZ_{\ijB} \ad_{\vecj} + \hc \rgh}$, in addition to the terms in Eq.~\eqref{eq_2DLGTDef}.
Nevertheless, we can consider two limits of our model, which correspond to the known results, derived in their seminal work.
In the limit where $\mu \rightarrow \infty$, our system has no particles, $N^a = 0$ and is thus a pure even \Zt gauge theory, with a confinement-deconfinement transition at a finite value $h/J \neq 0$ \cite{Fradkin1979}.
The deconfined phase has a \Zt topological order \cite{Borla2022}.
Two similar phases exist in the limit $\mu \rightarrow - \infty$, when the system has unit filling $n = N^{a} / (L_x \times L_y) = 1$, and corresponds to the odd \Zt LGT \cite{Borla2022}.
At finite bosonic parton filling, $0<n<1$, the phase diagram is not known and has only been explored for spinless fermionic matter \cite{Borla2022, Borla2024}.
There, interesting phases of matter were explored at half-filling $n = 0.5$, where at low values of the plaquette term $J \ll h,t$, a transition between a confined Mott state of dimers into a deconfined topological Dirac semi-metal was predicted \cite{Borla2022}.

\textit{Dynamically induced plaquette terms.---}
We start by simulating the (2+1)D \Zt LGT using DMRG \cite{White1992, Schollwoeck2011} on a cylinder of length $L_x$ and circumference $L_y$.
We use the MPS toolkit \textsc{SyTen} \cite{hubig:_syten_toolk, hubig17:_symmet_protec_tensor_network} to perform the DMRG calculations, see also Supplemental Material~\cite{SM} for details.

\begin{figure}[t]
\centering
\epsfig{file=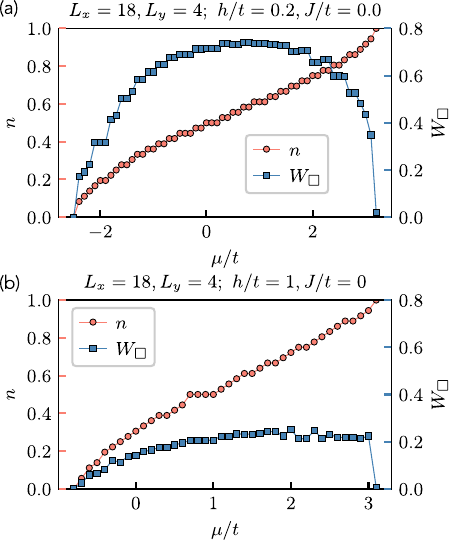}%, width=0.48\textwidth}
\caption{Lattice filling $n$ and the plaquette term $W_{\square}$ as a function of the chemical potential $\mu$ in the \Zt LGT Eq.~\eqref{eq_2DLGTDef} without the bare plaquette terms Eq.~\eqref{eq_Z2Plaquette}, $J = 0$. (a) Dynamical U(1) matter induces sizable finite value of the plaquette term $W_{\square}$ for $h = 0.2 t$. The value of $W_{\square}$ has a strong filling dependence and reaches its maximal value close to filling $n \approx 0.6$.
(b) The value of the plaquette term $W_{\square}$ decreases for stronger \Zt electric field $h = t$, but retains the filling dependence.
These results were obtained using DMRG on a cylinder of length $L_x = 18$ and circumference $L_y = 4$.
}
\label{FigTwo}
\end{figure}

We observe a finite value of the plaquette term 
\begin{equation}
    W_{\square} = \left \langle  \prod_{\ijB \in \square} \tauZ_{\ijB} \right \rangle,
    \label{eq_PlaquetteExpectation}
\end{equation}
for $J = 0$ and the finite value of the \Zt electric field term, $h$, across different filling $n$; see Fig.~\ref{FigTwo}.
For a low \Zt electric field term, $h/t = 0.2$, we observe a relatively high value of the plaquette operator.
Furthermore, the value of the plaquette term depends strongly on the filling $n$, which we control via the chemical potential term $\mu$; see Fig.~\ref{FigTwo}(a).
This indicates that the plaquette term is indeed generated by the U(1) matter.
The highest value of the plaquette term is found for parton filling around $n \approx 0.6$.

For a stronger \Zt electric field term $h /t = 1$, the plaquette values $W_\square$ are lower; see Fig.~\ref{FigTwo}(b).
In such a regime, the plaquette term values also have a filling dependence, and their maximum is at approximately $n \approx 2/3$.
For $h/t = 1$, we expect the mesons to be already tightly confined and thus less mobile, as they hop via a second-order process \cite{Borla2020, Kebric2021, Kebric2023}.

For $h/t = 0.2$, we expect that the mesons have already formed, as suggested by a finite value of the Fredenhagen-Marcu order parameter \cite{Fredenhagen1983, Fredenhagen1986, Marcu1986, Fredenhagen1988} obtained in this regime; see Supplemental Material~\cite{SM}.
However, string lengths can fluctuate considerably more than in the regime when $h/t = 1$, which results in their high mobility, thus inducing stronger plaquette terms.

These results agree with recent claims that dynamical matter can perturbatively induce plaquette terms \cite{Cobos2025}, even beyond special limits where this effect has been explicitly proven \cite{Homeier2021}.
Our results show that U(1) matter can enhance these effects and generate substantial plaquette values starting from $J=0$.

At higher electric field term $h /t =1$, we observe a plateau at half-filling $n = 1/2$ as a function of the chemical potential $\mu$; see Fig.~\ref{FigTwo}(b).
This indicates a gapped state.
However, a finite system size analysis is needed to determine if this gap corresponds to an incompressible meson state observed for fermionic matter \cite{Borla2022}.

%%%%%%%%%%%%%%%%%%%%%%%%%%%%%%%%%%%%%%%%%%%%%%%%%%%%%
%%%%%%%%%%%%%%%%%%%%%%%%%%%%%%%%%%%%%%%%%%%%%%%%%%%%%

\begin{figure}[t]
\centering
\epsfig{file=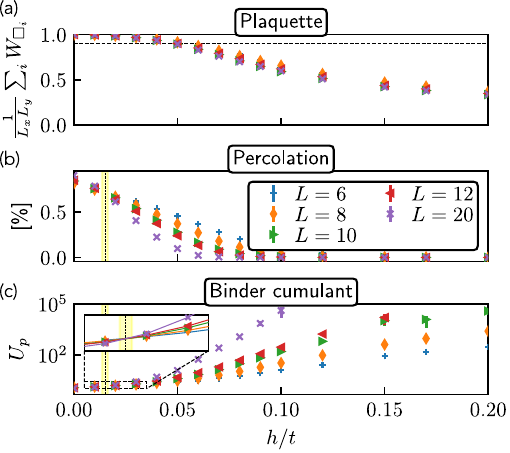}%, width=0.48\textwidth}
\caption{Plaquette expectation value and percolation-based quantities calculated with NQS, for the \Zt LGT coupled to dynamical matter Eq.~\eqref{eq_2DLGTDef} on a $L \times L$ square lattice at $20\%$ filling, $N^a= \{8,14,20,30,80\}$. (a) The average plaquette expectation value decreases as a function of $h/t$, while remaining approximately constant with respect to the system size. The dashed line shows the approximate plaquette expectation value at the phase transition in the perturbed toric code without matter. (b) The percolation probability and (c) the Binder cumulant of the percolation strength both indicate a phase transition near $h/t \approx 0.015$. The statistical errors are, in most cases, smaller than the respective marker.}
\label{Fig3}
\end{figure}

\textit{Large two-dimensional systems.---}
To investigate larger two-dimensional systems on a torus, we employ neural quantum states, which have been shown to efficiently represent a wide range of quantum states \cite{Carleo2017, Chen2024, Pfau2024,Szabo_2020,Chen_20202_sign,Lange_2024_RNN,Doeschl_2025,Kufel_2025,Moreno_2022,Roth2025,Lange_2024,Denis2025accurateneural,moss2025}, regardless of the choice of boundary conditions.
In our implementation, we enforce Gauss’s law in sampling (spins are defined in the $\hat{\tau}^x$-basis) and employ a deep neural network architecture tailored to the Hamiltonian, improving the representational power even for large systems \cite{Doeschl_2025_Correlations,Luo2021,Favoni2022, Apte2024}.
We use NQS to extend the numerical results discussed before by investigating $L\times L$ systems up to $L = 20$ for a filling of approximately $20 \%$ while tuning the electric field $h$; see Fig.~\ref{Fig3}.
Details about the NQS ansatz, the specific architecture used, and the parameter values are provided in the Supplemental Material~\cite{SM}.
 
Similar to the MPS calculations, we observe large plaquette expectation values $W_\square$ over a broad range of $h$ values; see Fig.~\ref{Fig3}(a).
This agrees with the argument that the increase in the coupling strength reduces the motion of matter and lowers the plaquette expectation value.
Notably, $W_\square$ remains essentially unchanged with increasing system size, even when evaluating large $L = 20$ systems, suggesting that this behavior is robust and likely persists in the thermodynamic limit.

Furthermore, the NQS calculations for the large system on the torus allow us to investigate the transition between the deconfined phase at $h = 0$ and the confined phase associated with $h\rightarrow \infty$.
To probe the confinement transition, we consider the percolation parameter and the Binder cumulant of the percolation strength \cite{Linsel2024, Linsel2025}.
Both indicators, the percolation parameter shown in Fig.~\ref{Fig3}(b), and the Binder cumulant of the percolation strength in Fig.~\ref{Fig3}(c), suggest a phase transition at $h/t \approx 0.015$.
Below we analyze the confinement-deconfinement transition and the plaquette value results in greater detail.
Numerical benchmarks between the DMRG and the NQS are presented in Supplemental Material~\cite{SM}. Furthermore, Supplemental Material~\cite{SM} contains additional data that provide a close up of the transition point in the binder cumulant and an analysis of the Fredenhagen-Marcu order parameter.

\textit{Plaquette value.---} For $h/t \rightarrow 0$, the plaquette expectation value approaches the maximum possible value, $W_\square = 1$; see Fig.~\ref{Fig3}(a).
Upon increasing $h/t$, $W_\square$ decreases and at $ h^\mathrm{int}/t \approx 0.05$, $W_\square$ intersects the approximate plaquette expectation value of the perturbed toric code at its phase transition point (found for parameters $h/J=0.328$ in the pure \Zt Ising gauge theory without matter \cite{Wu2012}, computed with quantum Monte Carlo package Paratoric \cite{Linsel2025_Paratoric} for $L=40$).
As there is no bare plaquette term in the Hamiltonian we simulate ($J=0$), our observation suggests a potential toric code-like phase transition around $h \approx h^{\mathrm{int}}$, explored further below.
For $h > h^\mathrm{int}$, the plaquette expectation value initially decreases rapidly, before gradually flattening out.

\textit{Confinement in the \Zt LGT.---} 
The numerical detection of a phase transition in a \Zt LGT is often done by investigating the Fredenhagen-Marcu order parameter \cite{Fredenhagen1983, Fredenhagen1986, Marcu1986, Fredenhagen1988, Gregor2011, Xu2025}.
Such an order parameter is constructed by dividing the expectation value of half a Wilson loop by a full loop.
Both expectation values decay exponentially for finite magnetic and electric field strengths \cite{Gregor2011}.
This makes it susceptible to numerical inaccuracies, particularly in variational calculations, which can yield misleading or unstable results \cite{Xu2025}.
Alternatively, as proposed in \cite{Linsel2024}, the percolation probability was shown to be a robust indicator for such phase transitions.
Unlike the Fredenhagen-Marcu order parameter, it does not suffer from comparable numerical instabilities.
In the following, we analyze the critical behavior of the system by discussing the percolation probability and its corresponding Binder cumulant $U_p(h,L) = \frac{\langle P^4(h,L)\rangle}{\langle P^2(h,L)\rangle^2}$, where $P(h,L)$ denotes the size of the percolating cluster \cite{Linsel2024}. 

Our results show a rapid decrease in the percolation probability as the confining field $h/t$ increases; see Fig.~\ref{Fig3}(b).
Specifically, at zero electric field strength, the percolation probability is at its maximum of about $80\%$ and decreases rapidly to $0\%$  with increasing electric field.

%%%%%%%%%%%%%%%%%%%%%%%%%%%%%%%%%%%%%%%%%%%%%%%%%%%%%
%%%%%%%%%%%%%%%%%%%%%%%%%%%%%%%%%%%%%%%%%%%%%%%%%%%%%

We observe that the slope of the percolation probability changes with system size, leading to a crossing point at $h/t \approx 0.015$.
This finite-size crossing is an indicator of a phase transition and is consistent with the expected behavior in a system with a confined-deconfined phase transition \cite{Linsel2024, Linsel2025}.

To support this conclusion, we analyze the Binder cumulant of the percolation strength.
A crossing of the Binder cumulant for different system sizes is a well-established indicator of criticality in many-body systems~\cite{Binder1981, Hasenbusch2008, Wu2012, Linsel2024}.
In Fig.~\ref{Fig3}(c), we observe such a crossing at $h/t \approx 0.015$, providing further evidence for a phase transition in this region.
In Supplemental Material~\cite{SM} we present a close up of the Binder cumulant crossing point, which shows reasonable convergence with the system size. Additionally, we evaluate the Fredenhagen–Marcu order parameter, which exhibits a change in its scaling behavior in the region where the Binder cumulant shows its crossing.

This indicates that the confinement-deconfinement transition occurs already for a very low \Zt electric field value $h$, and that mesons already become confined in the regime where we still observe large plaquette values.
In such regime, we furthermore observe evidence of meson condensation via extended-range off-diagonal meson correlations; see Supplemental Material~\cite{SM}.

\textit{Conclusion and outlook.---}
Quantum simulation of LGTs remains a challenge due to the necessity to engineer complicated multi-body interaction terms, in particular, to reach interesting deconfined topological phases.
Our results show that coupling U(1) matter to a \Zt LGT induces large values of the plaquette term, which is driven by mobile matter.
This reduces the requirement of engineering strong native plaquette terms in order to achieve interesting regimes in quantum simulation experiments.

We performed numerical calculations using MPS and NQS, both of which predict substantial plaquette term values on a cylinder and torus, for a system without native plaquette terms in the Hamiltonian.
With our NQS calculations, we were able to investigate significantly larger two-dimensional systems than accessible by the well established DMRG calculations on a cylinder; the latter were used in part to benchmark the NQS results.
We demonstrated that the NQS are a powerful new method for investigating LGTs with matter, thus opening a new avenue going beyond the capabilities of quantum Monte Carlo.

We applied our large-scale NQS simulations to start the exploration of the intriguing phase diagram of the \Zt LGT coupled to U(1) bosonic matter, focusing on a density of $n \approx 20 \%$.
We found strong numerical indications of a confinement-deconfinement transition around a small confining field $h \approx 0.015t$, possibly between a topological and non-topological phase.
To establish the complete phase diagram of the \Zt LGT coupled to U(1) matter, further detailed studies will be necessary, and NQS are a promising new platform which can be employed to explore significantly larger system sizes than previously accessible.
The effect of an emergent U(1) symmetry associated with total meson, rather than parton, number on the transition, and its relation to meson condensation is one of the possible future directions, which will provide an important cornerstone in our understanding of the \Zt LGTs with U(1) matter.
Our results thus pave the way for future quantum simulations of \Zt LGTs to explore these interesting phases of matter without the complicated implementation of native multi-body interactions.

\textit{Acknowledgments.---}
We thank Lukas Homeier and Simon Linsel for fruitful discussions.
This project has received funding from the European Research Council (ERC) under the European Union’s Horizon 2020 research and innovation programme (Grant Agreement no 948141) — ERC Starting Grant SimUcQuam.
This research was also supported by LMUexcellent, funded by the Federal Ministry of Education and Research (BMBF) and the Free State of Bavaria under the Excellence Strategy of the Federal Government and the Länder, and by the Deutsche Forschungsgemeinschaft (DFG, German Research Foundation) under Germany’s Excellence Strategy (EXC-2111 -- 390814868). 
U.B.~and J.C.H.~acknowledge funding by the Max Planck Society and the European Research Council (ERC) under the European Union’s Horizon Europe research and innovation program (Grant Agreement No.~101165667)—ERC Starting Grant QuSiGauge.

%%%%%%%%%%%%%%%%%%%%%%%%%%%%%%%%%%%%%
%\bibliography{U1MatterZ2LGT.bib}

\begin{thebibliography}{98}%
\makeatletter
\providecommand \@ifxundefined [1]{%
 \@ifx{#1\undefined}
}%
\providecommand \@ifnum [1]{%
 \ifnum #1\expandafter \@firstoftwo
 \else \expandafter \@secondoftwo
 \fi
}%
\providecommand \@ifx [1]{%
 \ifx #1\expandafter \@firstoftwo
 \else \expandafter \@secondoftwo
 \fi
}%
\providecommand \natexlab [1]{#1}%
\providecommand \enquote  [1]{``#1''}%
\providecommand \bibnamefont  [1]{#1}%
\providecommand \bibfnamefont [1]{#1}%
\providecommand \citenamefont [1]{#1}%
\providecommand \href@noop [0]{\@secondoftwo}%
\providecommand \href [0]{\begingroup \@sanitize@url \@href}%
\providecommand \@href[1]{\@@startlink{#1}\@@href}%
\providecommand \@@href[1]{\endgroup#1\@@endlink}%
\providecommand \@sanitize@url [0]{\catcode `\\12\catcode `\$12\catcode
  `\&12\catcode `\#12\catcode `\^12\catcode `\_12\catcode `\%12\relax}%
\providecommand \@@startlink[1]{}%
\providecommand \@@endlink[0]{}%
\providecommand \url  [0]{\begingroup\@sanitize@url \@url }%
\providecommand \@url [1]{\endgroup\@href {#1}{\urlprefix }}%
\providecommand \urlprefix  [0]{URL }%
\providecommand \Eprint [0]{\href }%
\providecommand \doibase [0]{http://dx.doi.org/}%
\providecommand \selectlanguage [0]{\@gobble}%
\providecommand \bibinfo  [0]{\@secondoftwo}%
\providecommand \bibfield  [0]{\@secondoftwo}%
\providecommand \translation [1]{[#1]}%
\providecommand \BibitemOpen [0]{}%
\providecommand \bibitemStop [0]{}%
\providecommand \bibitemNoStop [0]{.\EOS\space}%
\providecommand \EOS [0]{\spacefactor3000\relax}%
\providecommand \BibitemShut  [1]{\csname bibitem#1\endcsname}%
\let\auto@bib@innerbib\@empty
%</preamble>
\bibitem [{\citenamefont {Wilson}(1974)}]{Wilson1974}%
  \BibitemOpen
  \bibfield  {author} {\bibinfo {author} {\bibfnamefont {Kenneth~G.}\
  \bibnamefont {Wilson}},\ }\bibfield  {title} {\enquote {\bibinfo {title}
  {Confinement of quarks},}\ }\href {\doibase 10.1103/physrevd.10.2445}
  {\bibfield  {journal} {\bibinfo  {journal} {Physical Review D}\ }\textbf
  {\bibinfo {volume} {10}},\ \bibinfo {pages} {2445--2459} (\bibinfo {year}
  {1974})}\BibitemShut {NoStop}%
\bibitem [{\citenamefont {Kogut}(1979)}]{Kogut1979}%
  \BibitemOpen
  \bibfield  {author} {\bibinfo {author} {\bibfnamefont {John~B.}\ \bibnamefont
  {Kogut}},\ }\bibfield  {title} {\enquote {\bibinfo {title} {An introduction
  to lattice gauge theory and spin systems},}\ }\href {\doibase
  10.1103/revmodphys.51.659} {\bibfield  {journal} {\bibinfo  {journal}
  {Reviews of Modern Physics}\ }\textbf {\bibinfo {volume} {51}},\ \bibinfo
  {pages} {659--713} (\bibinfo {year} {1979})}\BibitemShut {NoStop}%
\bibitem [{\citenamefont {Fradkin}\ and\ \citenamefont
  {Shenker}(1979)}]{Fradkin1979}%
  \BibitemOpen
  \bibfield  {author} {\bibinfo {author} {\bibfnamefont {Eduardo}\ \bibnamefont
  {Fradkin}}\ and\ \bibinfo {author} {\bibfnamefont {Stephen~H.}\ \bibnamefont
  {Shenker}},\ }\bibfield  {title} {\enquote {\bibinfo {title} {Phase diagrams
  of lattice gauge theories with {H}iggs fields},}\ }\href {\doibase
  10.1103/physrevd.19.3682} {\bibfield  {journal} {\bibinfo  {journal}
  {Physical Review D}\ }\textbf {\bibinfo {volume} {19}},\ \bibinfo {pages}
  {3682--3697} (\bibinfo {year} {1979})}\BibitemShut {NoStop}%
\bibitem [{\citenamefont {Wen}(2007)}]{Wen2007}%
  \BibitemOpen
  \bibfield  {author} {\bibinfo {author} {\bibfnamefont {Xiao-Gang}\
  \bibnamefont {Wen}},\ }\href {\doibase
  10.1093/acprof:oso/9780199227259.001.0001} {\emph {\bibinfo {title} {Quantum
  Field Theory of Many-Body Systems}}}\ (\bibinfo  {publisher} {Oxford
  University Press},\ \bibinfo {year} {2007})\BibitemShut {NoStop}%
\bibitem [{\citenamefont {Sachdev}(2018)}]{Sachdev2018}%
  \BibitemOpen
  \bibfield  {author} {\bibinfo {author} {\bibfnamefont {Subir}\ \bibnamefont
  {Sachdev}},\ }\bibfield  {title} {\enquote {\bibinfo {title} {Topological
  order, emergent gauge fields, and {F}ermi surface reconstruction},}\ }\href
  {\doibase 10.1088/1361-6633/aae110} {\bibfield  {journal} {\bibinfo
  {journal} {Reports on Progress in Physics}\ }\textbf {\bibinfo {volume}
  {82}},\ \bibinfo {pages} {014001} (\bibinfo {year} {2018})}\BibitemShut
  {NoStop}%
\bibitem [{\citenamefont {Semeghini}\ \emph {et~al.}(2021)\citenamefont
  {Semeghini}, \citenamefont {Levine}, \citenamefont {Keesling}, \citenamefont
  {Ebadi}, \citenamefont {Wang}, \citenamefont {Bluvstein}, \citenamefont
  {Verresen}, \citenamefont {Pichler}, \citenamefont {Kalinowski},
  \citenamefont {Samajdar}, \citenamefont {Omran}, \citenamefont {Sachdev},
  \citenamefont {Vishwanath}, \citenamefont {Greiner}, \citenamefont
  {Vuletić},\ and\ \citenamefont {Lukin}}]{Semeghini2021}%
  \BibitemOpen
  \bibfield  {author} {\bibinfo {author} {\bibfnamefont {G.}~\bibnamefont
  {Semeghini}}, \bibinfo {author} {\bibfnamefont {H.}~\bibnamefont {Levine}},
  \bibinfo {author} {\bibfnamefont {A.}~\bibnamefont {Keesling}}, \bibinfo
  {author} {\bibfnamefont {S.}~\bibnamefont {Ebadi}}, \bibinfo {author}
  {\bibfnamefont {T.~T.}\ \bibnamefont {Wang}}, \bibinfo {author}
  {\bibfnamefont {D.}~\bibnamefont {Bluvstein}}, \bibinfo {author}
  {\bibfnamefont {R.}~\bibnamefont {Verresen}}, \bibinfo {author}
  {\bibfnamefont {H.}~\bibnamefont {Pichler}}, \bibinfo {author} {\bibfnamefont
  {M.}~\bibnamefont {Kalinowski}}, \bibinfo {author} {\bibfnamefont
  {R.}~\bibnamefont {Samajdar}}, \bibinfo {author} {\bibfnamefont
  {A.}~\bibnamefont {Omran}}, \bibinfo {author} {\bibfnamefont
  {S.}~\bibnamefont {Sachdev}}, \bibinfo {author} {\bibfnamefont
  {A.}~\bibnamefont {Vishwanath}}, \bibinfo {author} {\bibfnamefont
  {M.}~\bibnamefont {Greiner}}, \bibinfo {author} {\bibfnamefont
  {V.}~\bibnamefont {Vuletić}}, \ and\ \bibinfo {author} {\bibfnamefont
  {M.~D.}\ \bibnamefont {Lukin}},\ }\bibfield  {title} {\enquote {\bibinfo
  {title} {Probing topological spin liquids on a programmable quantum
  simulator},}\ }\href {\doibase 10.1126/science.abi8794} {\bibfield  {journal}
  {\bibinfo  {journal} {Science}\ }\textbf {\bibinfo {volume} {374}},\ \bibinfo
  {pages} {1242--1247} (\bibinfo {year} {2021})}\BibitemShut {NoStop}%
\bibitem [{\citenamefont {Senthil}\ and\ \citenamefont
  {Fisher}(2000)}]{Senthil2000}%
  \BibitemOpen
  \bibfield  {author} {\bibinfo {author} {\bibfnamefont {T.}~\bibnamefont
  {Senthil}}\ and\ \bibinfo {author} {\bibfnamefont {Matthew P.~A.}\
  \bibnamefont {Fisher}},\ }\bibfield  {title} {\enquote {\bibinfo {title}
  {{$Z_2$} gauge theory of electron fractionalization in strongly correlated
  systems},}\ }\href {\doibase 10.1103/physrevb.62.7850} {\bibfield  {journal}
  {\bibinfo  {journal} {Physical Review B}\ }\textbf {\bibinfo {volume} {62}},\
  \bibinfo {pages} {7850--7881} (\bibinfo {year} {2000})}\BibitemShut {NoStop}%
\bibitem [{\citenamefont {Sedgewick}\ \emph {et~al.}(2002)\citenamefont
  {Sedgewick}, \citenamefont {Scalapino},\ and\ \citenamefont
  {Sugar}}]{Sedgewick2002}%
  \BibitemOpen
  \bibfield  {author} {\bibinfo {author} {\bibfnamefont {R.}~\bibnamefont
  {Sedgewick}}, \bibinfo {author} {\bibfnamefont {D.}~\bibnamefont
  {Scalapino}}, \ and\ \bibinfo {author} {\bibfnamefont {R.}~\bibnamefont
  {Sugar}},\ }\bibfield  {title} {\enquote {\bibinfo {title} {Fractionalized
  phase in an {$XY$–$Z_2$} gauge model},}\ }\href {\doibase
  10.1103/physrevb.65.054508} {\bibfield  {journal} {\bibinfo  {journal}
  {Physical Review B}\ }\textbf {\bibinfo {volume} {65}},\ \bibinfo {pages}
  {054508} (\bibinfo {year} {2002})}\BibitemShut {NoStop}%
\bibitem [{\citenamefont {Gazit}\ \emph {et~al.}(2017)\citenamefont {Gazit},
  \citenamefont {Randeria},\ and\ \citenamefont {Vishwanath}}]{Gazit2017}%
  \BibitemOpen
  \bibfield  {author} {\bibinfo {author} {\bibfnamefont {Snir}\ \bibnamefont
  {Gazit}}, \bibinfo {author} {\bibfnamefont {Mohit}\ \bibnamefont {Randeria}},
  \ and\ \bibinfo {author} {\bibfnamefont {Ashvin}\ \bibnamefont
  {Vishwanath}},\ }\bibfield  {title} {\enquote {\bibinfo {title} {Emergent
  {D}irac fermions and broken symmetries in confined and deconfined phases of
  {$Z_2$} gauge theories},}\ }\href {\doibase 10.1038/nphys4028} {\bibfield
  {journal} {\bibinfo  {journal} {Nature Physics}\ }\textbf {\bibinfo {volume}
  {13}},\ \bibinfo {pages} {484--490} (\bibinfo {year} {2017})}\BibitemShut
  {NoStop}%
\bibitem [{\citenamefont {Demler}\ \emph {et~al.}(2002)\citenamefont {Demler},
  \citenamefont {Nayak}, \citenamefont {Kee}, \citenamefont {Kim},\ and\
  \citenamefont {Senthil}}]{Demler2002}%
  \BibitemOpen
  \bibfield  {author} {\bibinfo {author} {\bibfnamefont {Eugene}\ \bibnamefont
  {Demler}}, \bibinfo {author} {\bibfnamefont {Chetan}\ \bibnamefont {Nayak}},
  \bibinfo {author} {\bibfnamefont {Hae-Young}\ \bibnamefont {Kee}}, \bibinfo
  {author} {\bibfnamefont {Yong~Baek}\ \bibnamefont {Kim}}, \ and\ \bibinfo
  {author} {\bibfnamefont {T.}~\bibnamefont {Senthil}},\ }\bibfield  {title}
  {\enquote {\bibinfo {title} {Fractionalization patterns in strongly
  correlated electron systems: {S}pin-charge separation and beyond},}\ }\href
  {\doibase 10.1103/physrevb.65.155103} {\bibfield  {journal} {\bibinfo
  {journal} {Physical Review B}\ }\textbf {\bibinfo {volume} {65}},\ \bibinfo
  {pages} {155103} (\bibinfo {year} {2002})}\BibitemShut {NoStop}%
\bibitem [{\citenamefont {Kaul}\ \emph {et~al.}(2007)\citenamefont {Kaul},
  \citenamefont {Kim}, \citenamefont {Sachdev},\ and\ \citenamefont
  {Senthil}}]{Kaul2007}%
  \BibitemOpen
  \bibfield  {author} {\bibinfo {author} {\bibfnamefont {Ribhu~K.}\
  \bibnamefont {Kaul}}, \bibinfo {author} {\bibfnamefont {Yong~Baek}\
  \bibnamefont {Kim}}, \bibinfo {author} {\bibfnamefont {Subir}\ \bibnamefont
  {Sachdev}}, \ and\ \bibinfo {author} {\bibfnamefont {T.}~\bibnamefont
  {Senthil}},\ }\bibfield  {title} {\enquote {\bibinfo {title} {Algebraic
  charge liquids},}\ }\href {\doibase 10.1038/nphys790} {\bibfield  {journal}
  {\bibinfo  {journal} {Nature Physics}\ }\textbf {\bibinfo {volume} {4}},\
  \bibinfo {pages} {28--31} (\bibinfo {year} {2007})}\BibitemShut {NoStop}%
\bibitem [{\citenamefont {Kitaev}(2003)}]{Kitaev2003}%
  \BibitemOpen
  \bibfield  {author} {\bibinfo {author} {\bibfnamefont {A.Yu.}\ \bibnamefont
  {Kitaev}},\ }\bibfield  {title} {\enquote {\bibinfo {title} {Fault-tolerant
  quantum computation by anyons},}\ }\href {\doibase
  10.1016/s0003-4916(02)00018-0} {\bibfield  {journal} {\bibinfo  {journal}
  {Annals of Physics}\ }\textbf {\bibinfo {volume} {303}},\ \bibinfo {pages}
  {2--30} (\bibinfo {year} {2003})}\BibitemShut {NoStop}%
\bibitem [{\citenamefont {Carmen~Bañuls}\ and\ \citenamefont
  {Cichy}(2020)}]{CarmenBanuls2020}%
  \BibitemOpen
  \bibfield  {author} {\bibinfo {author} {\bibfnamefont {Mari}\ \bibnamefont
  {Carmen~Bañuls}}\ and\ \bibinfo {author} {\bibfnamefont {Krzysztof}\
  \bibnamefont {Cichy}},\ }\bibfield  {title} {\enquote {\bibinfo {title}
  {Review on novel methods for lattice gauge theories},}\ }\href {\doibase
  10.1088/1361-6633/ab6311} {\bibfield  {journal} {\bibinfo  {journal} {Reports
  on Progress in Physics}\ }\textbf {\bibinfo {volume} {83}},\ \bibinfo {pages}
  {024401} (\bibinfo {year} {2020})}\BibitemShut {NoStop}%
\bibitem [{\citenamefont {Greensite}(2011)}]{Greensite2011}%
  \BibitemOpen
  \bibfield  {author} {\bibinfo {author} {\bibfnamefont {Jeff}\ \bibnamefont
  {Greensite}},\ }\href {\doibase 10.1007/978-3-642-14382-3} {\emph {\bibinfo
  {title} {An Introduction to the Confinement Problem}}}\ (\bibinfo
  {publisher} {Springer Berlin Heidelberg},\ \bibinfo {year}
  {2011})\BibitemShut {NoStop}%
\bibitem [{\citenamefont {Schollwöck}(2011)}]{Schollwoeck2011}%
  \BibitemOpen
  \bibfield  {author} {\bibinfo {author} {\bibfnamefont {Ulrich}\ \bibnamefont
  {Schollwöck}},\ }\bibfield  {title} {\enquote {\bibinfo {title} {The
  density-matrix renormalization group in the age of matrix product states},}\
  }\href {\doibase 10.1016/j.aop.2010.09.012} {\bibfield  {journal} {\bibinfo
  {journal} {Annals of Physics}\ }\textbf {\bibinfo {volume} {326}},\ \bibinfo
  {pages} {96--192} (\bibinfo {year} {2011})}\BibitemShut {NoStop}%
\bibitem [{\citenamefont {White}(1992)}]{White1992}%
  \BibitemOpen
  \bibfield  {author} {\bibinfo {author} {\bibfnamefont {Steven~R.}\
  \bibnamefont {White}},\ }\bibfield  {title} {\enquote {\bibinfo {title}
  {Density matrix formulation for quantum renormalization groups},}\ }\href
  {\doibase 10.1103/physrevlett.69.2863} {\bibfield  {journal} {\bibinfo
  {journal} {Physical Review Letters}\ }\textbf {\bibinfo {volume} {69}},\
  \bibinfo {pages} {2863--2866} (\bibinfo {year} {1992})}\BibitemShut {NoStop}%
\bibitem [{\citenamefont {Troyer}\ and\ \citenamefont
  {Wiese}(2005)}]{Troyer2005}%
  \BibitemOpen
  \bibfield  {author} {\bibinfo {author} {\bibfnamefont {Matthias}\
  \bibnamefont {Troyer}}\ and\ \bibinfo {author} {\bibfnamefont {Uwe-Jens}\
  \bibnamefont {Wiese}},\ }\bibfield  {title} {\enquote {\bibinfo {title}
  {Computational complexity and fundamental limitations to fermionic quantum
  {M}onte {C}arlo simulations},}\ }\href {\doibase
  10.1103/physrevlett.94.170201} {\bibfield  {journal} {\bibinfo  {journal}
  {Physical Review Letters}\ }\textbf {\bibinfo {volume} {94}},\ \bibinfo
  {pages} {170201} (\bibinfo {year} {2005})}\BibitemShut {NoStop}%
\bibitem [{\citenamefont {Assaad}\ and\ \citenamefont
  {Grover}(2016)}]{Assaad2016}%
  \BibitemOpen
  \bibfield  {author} {\bibinfo {author} {\bibfnamefont {F.~F.}\ \bibnamefont
  {Assaad}}\ and\ \bibinfo {author} {\bibfnamefont {Tarun}\ \bibnamefont
  {Grover}},\ }\bibfield  {title} {\enquote {\bibinfo {title} {Simple fermionic
  model of deconfined phases and phase transitions},}\ }\href {\doibase
  10.1103/physrevx.6.041049} {\bibfield  {journal} {\bibinfo  {journal}
  {Physical Review X}\ }\textbf {\bibinfo {volume} {6}},\ \bibinfo {pages}
  {041049} (\bibinfo {year} {2016})}\BibitemShut {NoStop}%
\bibitem [{\citenamefont {Bañuls}\ \emph {et~al.}(2020)\citenamefont
  {Bañuls}, \citenamefont {Blatt}, \citenamefont {Catani}, \citenamefont
  {Celi}, \citenamefont {Cirac}, \citenamefont {Dalmonte}, \citenamefont
  {Fallani}, \citenamefont {Jansen}, \citenamefont {Lewenstein}, \citenamefont
  {Montangero} \emph {et~al.}}]{Banuls2020}%
  \BibitemOpen
  \bibfield  {author} {\bibinfo {author} {\bibfnamefont {Mari~Carmen}\
  \bibnamefont {Bañuls}}, \bibinfo {author} {\bibfnamefont {Rainer}\
  \bibnamefont {Blatt}}, \bibinfo {author} {\bibfnamefont {Jacopo}\
  \bibnamefont {Catani}}, \bibinfo {author} {\bibfnamefont {Alessio}\
  \bibnamefont {Celi}}, \bibinfo {author} {\bibfnamefont {Juan~Ignacio}\
  \bibnamefont {Cirac}}, \bibinfo {author} {\bibfnamefont {Marcello}\
  \bibnamefont {Dalmonte}}, \bibinfo {author} {\bibfnamefont {Leonardo}\
  \bibnamefont {Fallani}}, \bibinfo {author} {\bibfnamefont {Karl}\
  \bibnamefont {Jansen}}, \bibinfo {author} {\bibfnamefont {Maciej}\
  \bibnamefont {Lewenstein}}, \bibinfo {author} {\bibfnamefont {Simone}\
  \bibnamefont {Montangero}},  \emph {et~al.},\ }\bibfield  {title} {\enquote
  {\bibinfo {title} {Simulating lattice gauge theories within quantum
  technologies},}\ }\href {\doibase 10.1140/epjd/e2020-100571-8} {\bibfield
  {journal} {\bibinfo  {journal} {The European Physical Journal D}\ }\textbf
  {\bibinfo {volume} {74}} (\bibinfo {year} {2020}),\
  10.1140/epjd/e2020-100571-8}\BibitemShut {NoStop}%
\bibitem [{\citenamefont {Magnifico}\ \emph {et~al.}(2025)\citenamefont
  {Magnifico}, \citenamefont {Cataldi}, \citenamefont {Rigobello},
  \citenamefont {Majcen}, \citenamefont {Jaschke}, \citenamefont {Silvi},\ and\
  \citenamefont {Montangero}}]{Magnifico2025}%
  \BibitemOpen
  \bibfield  {author} {\bibinfo {author} {\bibfnamefont {Giuseppe}\
  \bibnamefont {Magnifico}}, \bibinfo {author} {\bibfnamefont {Giovanni}\
  \bibnamefont {Cataldi}}, \bibinfo {author} {\bibfnamefont {Marco}\
  \bibnamefont {Rigobello}}, \bibinfo {author} {\bibfnamefont {Peter}\
  \bibnamefont {Majcen}}, \bibinfo {author} {\bibfnamefont {Daniel}\
  \bibnamefont {Jaschke}}, \bibinfo {author} {\bibfnamefont {Pietro}\
  \bibnamefont {Silvi}}, \ and\ \bibinfo {author} {\bibfnamefont {Simone}\
  \bibnamefont {Montangero}},\ }\bibfield  {title} {\enquote {\bibinfo {title}
  {Tensor networks for lattice gauge theories beyond one dimension},}\ }\href
  {\doibase 10.1038/s42005-025-02125-x} {\bibfield  {journal} {\bibinfo
  {journal} {Communications Physics}\ }\textbf {\bibinfo {volume} {8}}
  (\bibinfo {year} {2025}),\ 10.1038/s42005-025-02125-x}\BibitemShut {NoStop}%
\bibitem [{\citenamefont {Carleo}\ and\ \citenamefont
  {Troyer}(2017)}]{Carleo2017}%
  \BibitemOpen
  \bibfield  {author} {\bibinfo {author} {\bibfnamefont {Giuseppe}\
  \bibnamefont {Carleo}}\ and\ \bibinfo {author} {\bibfnamefont {Matthias}\
  \bibnamefont {Troyer}},\ }\bibfield  {title} {\enquote {\bibinfo {title}
  {Solving the quantum many-body problem with artificial neural networks},}\
  }\href {\doibase 10.1126/science.aag2302} {\bibfield  {journal} {\bibinfo
  {journal} {Science}\ }\textbf {\bibinfo {volume} {355}},\ \bibinfo {pages}
  {602--606} (\bibinfo {year} {2017})}\BibitemShut {NoStop}%
\bibitem [{\citenamefont {Chen}\ and\ \citenamefont {Heyl}(2024)}]{Chen2024}%
  \BibitemOpen
  \bibfield  {author} {\bibinfo {author} {\bibfnamefont {Ao}~\bibnamefont
  {Chen}}\ and\ \bibinfo {author} {\bibfnamefont {Markus}\ \bibnamefont
  {Heyl}},\ }\bibfield  {title} {\enquote {\bibinfo {title} {Empowering deep
  neural quantum states through efficient optimization},}\ }\href {\doibase
  10.1038/s41567-024-02566-1} {\bibfield  {journal} {\bibinfo  {journal}
  {Nature Physics}\ }\textbf {\bibinfo {volume} {20}},\ \bibinfo {pages}
  {1476--1481} (\bibinfo {year} {2024})}\BibitemShut {NoStop}%
\bibitem [{\citenamefont {Pfau}\ \emph {et~al.}(2024)\citenamefont {Pfau},
  \citenamefont {Axelrod}, \citenamefont {Sutterud}, \citenamefont {von
  Glehn},\ and\ \citenamefont {Spencer}}]{Pfau2024}%
  \BibitemOpen
  \bibfield  {author} {\bibinfo {author} {\bibfnamefont {David}\ \bibnamefont
  {Pfau}}, \bibinfo {author} {\bibfnamefont {Simon}\ \bibnamefont {Axelrod}},
  \bibinfo {author} {\bibfnamefont {Halvard}\ \bibnamefont {Sutterud}},
  \bibinfo {author} {\bibfnamefont {Ingrid}\ \bibnamefont {von Glehn}}, \ and\
  \bibinfo {author} {\bibfnamefont {James~S.}\ \bibnamefont {Spencer}},\
  }\bibfield  {title} {\enquote {\bibinfo {title} {Accurate computation of
  quantum excited states with neural networks},}\ }\href {\doibase
  10.1126/science.adn0137} {\bibfield  {journal} {\bibinfo  {journal}
  {Science}\ }\textbf {\bibinfo {volume} {385}} (\bibinfo {year} {2024}),\
  10.1126/science.adn0137}\BibitemShut {NoStop}%
\bibitem [{\citenamefont {Szab\'o}\ and\ \citenamefont
  {Castelnovo}(2020)}]{Szabo_2020}%
  \BibitemOpen
  \bibfield  {author} {\bibinfo {author} {\bibfnamefont {Attila}\ \bibnamefont
  {Szab\'o}}\ and\ \bibinfo {author} {\bibfnamefont {Claudio}\ \bibnamefont
  {Castelnovo}},\ }\bibfield  {title} {\enquote {\bibinfo {title} {Neural
  network wave functions and the sign problem},}\ }\href {\doibase
  10.1103/PhysRevResearch.2.033075} {\bibfield  {journal} {\bibinfo  {journal}
  {Phys. Rev. Res.}\ }\textbf {\bibinfo {volume} {2}},\ \bibinfo {pages}
  {033075} (\bibinfo {year} {2020})}\BibitemShut {NoStop}%
\bibitem [{\citenamefont {Chen}\ \emph {et~al.}(2022)\citenamefont {Chen},
  \citenamefont {Choo}, \citenamefont {Astrakhantsev},\ and\ \citenamefont
  {Neupert}}]{Chen_20202_sign}%
  \BibitemOpen
  \bibfield  {author} {\bibinfo {author} {\bibfnamefont {Ao}~\bibnamefont
  {Chen}}, \bibinfo {author} {\bibfnamefont {Kenny}\ \bibnamefont {Choo}},
  \bibinfo {author} {\bibfnamefont {Nikita}\ \bibnamefont {Astrakhantsev}}, \
  and\ \bibinfo {author} {\bibfnamefont {Titus}\ \bibnamefont {Neupert}},\
  }\bibfield  {title} {\enquote {\bibinfo {title} {Neural network evolution
  strategy for solving quantum sign structures},}\ }\href {\doibase
  10.1103/PhysRevResearch.4.L022026} {\bibfield  {journal} {\bibinfo  {journal}
  {Phys. Rev. Res.}\ }\textbf {\bibinfo {volume} {4}},\ \bibinfo {pages}
  {L022026} (\bibinfo {year} {2022})}\BibitemShut {NoStop}%
\bibitem [{\citenamefont {Lange}\ \emph
  {et~al.}(2024{\natexlab{a}})\citenamefont {Lange}, \citenamefont
  {D{\"o}schl}, \citenamefont {Carrasquilla},\ and\ \citenamefont
  {Bohrdt}}]{Lange_2024_RNN}%
  \BibitemOpen
  \bibfield  {author} {\bibinfo {author} {\bibfnamefont {Hannah}\ \bibnamefont
  {Lange}}, \bibinfo {author} {\bibfnamefont {Fabian}\ \bibnamefont
  {D{\"o}schl}}, \bibinfo {author} {\bibfnamefont {Juan}\ \bibnamefont
  {Carrasquilla}}, \ and\ \bibinfo {author} {\bibfnamefont {Annabelle}\
  \bibnamefont {Bohrdt}},\ }\bibfield  {title} {\enquote {\bibinfo {title}
  {Neural network approach to quasiparticle dispersions in doped
  antiferromagnets},}\ }\href {\doibase 10.1038/s42005-024-01678-7} {\bibfield
  {journal} {\bibinfo  {journal} {Communications Physics}\ }\textbf {\bibinfo
  {volume} {7}},\ \bibinfo {pages} {187} (\bibinfo {year}
  {2024}{\natexlab{a}})}\BibitemShut {NoStop}%
\bibitem [{\citenamefont {D\"oschl}\ \emph {et~al.}(2025)\citenamefont
  {D\"oschl}, \citenamefont {Palm}, \citenamefont {Lange}, \citenamefont
  {Grusdt},\ and\ \citenamefont {Bohrdt}}]{Doeschl_2025}%
  \BibitemOpen
  \bibfield  {author} {\bibinfo {author} {\bibfnamefont {Fabian}\ \bibnamefont
  {D\"oschl}}, \bibinfo {author} {\bibfnamefont {Felix~A.}\ \bibnamefont
  {Palm}}, \bibinfo {author} {\bibfnamefont {Hannah}\ \bibnamefont {Lange}},
  \bibinfo {author} {\bibfnamefont {Fabian}\ \bibnamefont {Grusdt}}, \ and\
  \bibinfo {author} {\bibfnamefont {Annabelle}\ \bibnamefont {Bohrdt}},\
  }\bibfield  {title} {\enquote {\bibinfo {title} {Neural network quantum
  states for the interacting {H}ofstadter model with higher local occupations
  and long-range interactions},}\ }\href {\doibase 10.1103/PhysRevB.111.045408}
  {\bibfield  {journal} {\bibinfo  {journal} {Phys. Rev. B}\ }\textbf {\bibinfo
  {volume} {111}},\ \bibinfo {pages} {045408} (\bibinfo {year}
  {2025})}\BibitemShut {NoStop}%
\bibitem [{\citenamefont {Kufel}\ \emph {et~al.}(2025)\citenamefont {Kufel},
  \citenamefont {Kemp}, \citenamefont {Vu}, \citenamefont {Linsel},
  \citenamefont {Laumann},\ and\ \citenamefont {Yao}}]{Kufel_2025}%
  \BibitemOpen
  \bibfield  {author} {\bibinfo {author} {\bibfnamefont {Dominik~S.}\
  \bibnamefont {Kufel}}, \bibinfo {author} {\bibfnamefont {Jack}\ \bibnamefont
  {Kemp}}, \bibinfo {author} {\bibfnamefont {DinhDuy}\ \bibnamefont {Vu}},
  \bibinfo {author} {\bibfnamefont {Simon~M.}\ \bibnamefont {Linsel}}, \bibinfo
  {author} {\bibfnamefont {Chris~R.}\ \bibnamefont {Laumann}}, \ and\ \bibinfo
  {author} {\bibfnamefont {Norman~Y.}\ \bibnamefont {Yao}},\ }\bibfield
  {title} {\enquote {\bibinfo {title} {Approximately symmetric neural networks
  for quantum spin liquids},}\ }\href {\doibase 10.1103/pgnx-11ph} {\bibfield
  {journal} {\bibinfo  {journal} {Phys. Rev. Lett.}\ }\textbf {\bibinfo
  {volume} {135}},\ \bibinfo {pages} {056702} (\bibinfo {year}
  {2025})}\BibitemShut {NoStop}%
\bibitem [{\citenamefont {Robledo~Moreno}\ \emph {et~al.}(2022)\citenamefont
  {Robledo~Moreno}, \citenamefont {Carleo}, \citenamefont {Georges},\ and\
  \citenamefont {Stokes}}]{Moreno_2022}%
  \BibitemOpen
  \bibfield  {author} {\bibinfo {author} {\bibfnamefont {Javier}\ \bibnamefont
  {Robledo~Moreno}}, \bibinfo {author} {\bibfnamefont {Giuseppe}\ \bibnamefont
  {Carleo}}, \bibinfo {author} {\bibfnamefont {Antoine}\ \bibnamefont
  {Georges}}, \ and\ \bibinfo {author} {\bibfnamefont {James}\ \bibnamefont
  {Stokes}},\ }\bibfield  {title} {\enquote {\bibinfo {title} {Fermionic wave
  functions from neural-network constrained hidden states},}\ }\href {\doibase
  10.1073/pnas.2122059119} {\bibfield  {journal} {\bibinfo  {journal}
  {Proceedings of the National Academy of Sciences}\ }\textbf {\bibinfo
  {volume} {119}} (\bibinfo {year} {2022}),\
  10.1073/pnas.2122059119}\BibitemShut {NoStop}%
\bibitem [{\citenamefont {Viteritti}\ \emph {et~al.}(2023)\citenamefont
  {Viteritti}, \citenamefont {Rende},\ and\ \citenamefont
  {Becca}}]{Viteritti2023}%
  \BibitemOpen
  \bibfield  {author} {\bibinfo {author} {\bibfnamefont {Luciano~Loris}\
  \bibnamefont {Viteritti}}, \bibinfo {author} {\bibfnamefont {Riccardo}\
  \bibnamefont {Rende}}, \ and\ \bibinfo {author} {\bibfnamefont {Federico}\
  \bibnamefont {Becca}},\ }\bibfield  {title} {\enquote {\bibinfo {title}
  {Transformer {V}ariational {W}ave {F}unctions for {F}rustrated {Q}uantum
  {S}pin {S}ystems},}\ }\href {\doibase 10.1103/physrevlett.130.236401}
  {\bibfield  {journal} {\bibinfo  {journal} {Physical Review Letters}\
  }\textbf {\bibinfo {volume} {130}},\ \bibinfo {pages} {236401} (\bibinfo
  {year} {2023})}\BibitemShut {NoStop}%
\bibitem [{\citenamefont {Roth}\ \emph {et~al.}(2025)\citenamefont {Roth},
  \citenamefont {Chen}, \citenamefont {Sengupta},\ and\ \citenamefont
  {Georges}}]{Roth2025}%
  \BibitemOpen
  \bibfield  {author} {\bibinfo {author} {\bibfnamefont {Christopher}\
  \bibnamefont {Roth}}, \bibinfo {author} {\bibfnamefont {Ao}~\bibnamefont
  {Chen}}, \bibinfo {author} {\bibfnamefont {Anirvan}\ \bibnamefont
  {Sengupta}}, \ and\ \bibinfo {author} {\bibfnamefont {Antoine}\ \bibnamefont
  {Georges}},\ }\href {https://arxiv.org/abs/2511.07566} {\enquote {\bibinfo
  {title} {Superconductivity in the two-dimensional {H}ubbard model revealed by
  neural quantum states},}\ } (\bibinfo {year} {2025}),\ \Eprint
  {http://arxiv.org/abs/2511.07566} {arXiv:2511.07566 [cond-mat.supr-con]}
  \BibitemShut {NoStop}%
\bibitem [{\citenamefont {Lange}\ \emph
  {et~al.}(2024{\natexlab{b}})\citenamefont {Lange}, \citenamefont {Van~de
  Walle}, \citenamefont {Abedinnia},\ and\ \citenamefont
  {Bohrdt}}]{Lange_2024}%
  \BibitemOpen
  \bibfield  {author} {\bibinfo {author} {\bibfnamefont {Hannah}\ \bibnamefont
  {Lange}}, \bibinfo {author} {\bibfnamefont {Anka}\ \bibnamefont {Van~de
  Walle}}, \bibinfo {author} {\bibfnamefont {Atiye}\ \bibnamefont {Abedinnia}},
  \ and\ \bibinfo {author} {\bibfnamefont {Annabelle}\ \bibnamefont {Bohrdt}},\
  }\bibfield  {title} {\enquote {\bibinfo {title} {From architectures to
  applications: a review of neural quantum states},}\ }\href {\doibase
  10.1088/2058-9565/ad7168} {\bibfield  {journal} {\bibinfo  {journal} {Quantum
  Science and Technology}\ }\textbf {\bibinfo {volume} {9}},\ \bibinfo {pages}
  {040501} (\bibinfo {year} {2024}{\natexlab{b}})}\BibitemShut {NoStop}%
\bibitem [{\citenamefont {Denis}\ and\ \citenamefont
  {Carleo}(2025)}]{Denis2025accurateneural}%
  \BibitemOpen
  \bibfield  {author} {\bibinfo {author} {\bibfnamefont {Zakari}\ \bibnamefont
  {Denis}}\ and\ \bibinfo {author} {\bibfnamefont {Giuseppe}\ \bibnamefont
  {Carleo}},\ }\bibfield  {title} {\enquote {\bibinfo {title} {Accurate neural
  quantum states for interacting lattice bosons},}\ }\href {\doibase
  10.22331/q-2025-06-17-1772} {\bibfield  {journal} {\bibinfo  {journal}
  {{Quantum}}\ }\textbf {\bibinfo {volume} {9}},\ \bibinfo {pages} {1772}
  (\bibinfo {year} {2025})}\BibitemShut {NoStop}%
\bibitem [{\citenamefont {Moss}\ \emph {et~al.}(2025)\citenamefont {Moss},
  \citenamefont {Wiersema}, \citenamefont {Hibat-Allah}, \citenamefont
  {Carrasquilla},\ and\ \citenamefont {Melko}}]{moss2025}%
  \BibitemOpen
  \bibfield  {author} {\bibinfo {author} {\bibfnamefont {M.~Schuyler}\
  \bibnamefont {Moss}}, \bibinfo {author} {\bibfnamefont {Roeland}\
  \bibnamefont {Wiersema}}, \bibinfo {author} {\bibfnamefont {Mohamed}\
  \bibnamefont {Hibat-Allah}}, \bibinfo {author} {\bibfnamefont {Juan}\
  \bibnamefont {Carrasquilla}}, \ and\ \bibinfo {author} {\bibfnamefont
  {Roger~G.}\ \bibnamefont {Melko}},\ }\href {https://arxiv.org/abs/2502.17144}
  {\enquote {\bibinfo {title} {Leveraging recurrence in neural network
  wavefunctions for large-scale simulations of {H}eisenberg antiferromagnets:
  the square lattice},}\ } (\bibinfo {year} {2025}),\ \Eprint
  {http://arxiv.org/abs/2502.17144} {arXiv:2502.17144 [cond-mat.str-el]}
  \BibitemShut {NoStop}%
\bibitem [{\citenamefont {Schmitt}\ and\ \citenamefont
  {Heyl}(2020)}]{Schmitt_2020}%
  \BibitemOpen
  \bibfield  {author} {\bibinfo {author} {\bibfnamefont {Markus}\ \bibnamefont
  {Schmitt}}\ and\ \bibinfo {author} {\bibfnamefont {Markus}\ \bibnamefont
  {Heyl}},\ }\bibfield  {title} {\enquote {\bibinfo {title} {Quantum many-body
  dynamics in two dimensions with artificial neural networks},}\ }\href
  {\doibase 10.1103/PhysRevLett.125.100503} {\bibfield  {journal} {\bibinfo
  {journal} {Phys. Rev. Lett.}\ }\textbf {\bibinfo {volume} {125}},\ \bibinfo
  {pages} {100503} (\bibinfo {year} {2020})}\BibitemShut {NoStop}%
\bibitem [{\citenamefont {Gutiérrez}\ and\ \citenamefont
  {Mendl}(2022)}]{Guti_rrez_2022}%
  \BibitemOpen
  \bibfield  {author} {\bibinfo {author} {\bibfnamefont {Irene~López}\
  \bibnamefont {Gutiérrez}}\ and\ \bibinfo {author} {\bibfnamefont
  {Christian~B.}\ \bibnamefont {Mendl}},\ }\bibfield  {title} {\enquote
  {\bibinfo {title} {Real time evolution with neural-network quantum states},}\
  }\href {\doibase 10.22331/q-2022-01-20-627} {\bibfield  {journal} {\bibinfo
  {journal} {Quantum}\ }\textbf {\bibinfo {volume} {6}},\ \bibinfo {pages}
  {627} (\bibinfo {year} {2022})}\BibitemShut {NoStop}%
\bibitem [{\citenamefont {Donatella}\ \emph {et~al.}(2023)\citenamefont
  {Donatella}, \citenamefont {Denis}, \citenamefont {Le~Boit\'e},\ and\
  \citenamefont {Ciuti}}]{Donatella_2023}%
  \BibitemOpen
  \bibfield  {author} {\bibinfo {author} {\bibfnamefont {Kaelan}\ \bibnamefont
  {Donatella}}, \bibinfo {author} {\bibfnamefont {Zakari}\ \bibnamefont
  {Denis}}, \bibinfo {author} {\bibfnamefont {Alexandre}\ \bibnamefont
  {Le~Boit\'e}}, \ and\ \bibinfo {author} {\bibfnamefont {Cristiano}\
  \bibnamefont {Ciuti}},\ }\bibfield  {title} {\enquote {\bibinfo {title}
  {Dynamics with autoregressive neural quantum states: {A}pplication to
  critical quench dynamics},}\ }\href {\doibase 10.1103/PhysRevA.108.022210}
  {\bibfield  {journal} {\bibinfo  {journal} {Phys. Rev. A}\ }\textbf {\bibinfo
  {volume} {108}},\ \bibinfo {pages} {022210} (\bibinfo {year}
  {2023})}\BibitemShut {NoStop}%
\bibitem [{\citenamefont {de~Walle}\ \emph {et~al.}(2024)\citenamefont
  {de~Walle}, \citenamefont {Schmitt},\ and\ \citenamefont
  {Bohrdt}}]{Walle_2024}%
  \BibitemOpen
  \bibfield  {author} {\bibinfo {author} {\bibfnamefont {Anka~Van}\
  \bibnamefont {de~Walle}}, \bibinfo {author} {\bibfnamefont {Markus}\
  \bibnamefont {Schmitt}}, \ and\ \bibinfo {author} {\bibfnamefont {Annabelle}\
  \bibnamefont {Bohrdt}},\ }\href {https://arxiv.org/abs/2412.11830} {\enquote
  {\bibinfo {title} {Many-body dynamics with explicitly time-dependent neural
  quantum states},}\ } (\bibinfo {year} {2024}),\ \Eprint
  {http://arxiv.org/abs/2412.11830} {arXiv:2412.11830 [quant-ph]} \BibitemShut
  {NoStop}%
\bibitem [{\citenamefont {Sinibaldi}\ \emph {et~al.}(2025)\citenamefont
  {Sinibaldi}, \citenamefont {Hendry}, \citenamefont {Vicentini},\ and\
  \citenamefont {Carleo}}]{sinibaldi_2025}%
  \BibitemOpen
  \bibfield  {author} {\bibinfo {author} {\bibfnamefont {Alessandro}\
  \bibnamefont {Sinibaldi}}, \bibinfo {author} {\bibfnamefont {Douglas}\
  \bibnamefont {Hendry}}, \bibinfo {author} {\bibfnamefont {Filippo}\
  \bibnamefont {Vicentini}}, \ and\ \bibinfo {author} {\bibfnamefont
  {Giuseppe}\ \bibnamefont {Carleo}},\ }\href
  {https://arxiv.org/abs/2412.11778} {\enquote {\bibinfo {title}
  {Time-dependent neural {G}alerkin method for quantum dynamics},}\ } (\bibinfo
  {year} {2025}),\ \Eprint {http://arxiv.org/abs/2412.11778} {arXiv:2412.11778
  [quant-ph]} \BibitemShut {NoStop}%
\bibitem [{\citenamefont {Schmitt}\ and\ \citenamefont
  {Heyl}(2025)}]{Schmitt_2025}%
  \BibitemOpen
  \bibfield  {author} {\bibinfo {author} {\bibfnamefont {Markus}\ \bibnamefont
  {Schmitt}}\ and\ \bibinfo {author} {\bibfnamefont {Markus}\ \bibnamefont
  {Heyl}},\ }\href {https://arxiv.org/abs/2506.03124} {\enquote {\bibinfo
  {title} {Simulating dynamics of correlated matter with neural quantum
  states},}\ } (\bibinfo {year} {2025}),\ \Eprint
  {http://arxiv.org/abs/2506.03124} {arXiv:2506.03124 [quant-ph]} \BibitemShut
  {NoStop}%
\bibitem [{\citenamefont {Gao}\ and\ \citenamefont {Duan}(2017)}]{Gao_2017}%
  \BibitemOpen
  \bibfield  {author} {\bibinfo {author} {\bibfnamefont {Xun}\ \bibnamefont
  {Gao}}\ and\ \bibinfo {author} {\bibfnamefont {Lu-Ming}\ \bibnamefont
  {Duan}},\ }\bibfield  {title} {\enquote {\bibinfo {title} {Efficient
  representation of quantum many-body states with deep neural networks},}\
  }\href {\doibase 10.1038/s41467-017-00705-2} {\bibfield  {journal} {\bibinfo
  {journal} {Nature Communications}\ }\textbf {\bibinfo {volume} {8}},\
  \bibinfo {pages} {662} (\bibinfo {year} {2017})}\BibitemShut {NoStop}%
\bibitem [{\citenamefont {Deng}\ \emph {et~al.}(2017)\citenamefont {Deng},
  \citenamefont {Li},\ and\ \citenamefont {Das~Sarma}}]{Deng_Quantum_Ent_2017}%
  \BibitemOpen
  \bibfield  {author} {\bibinfo {author} {\bibfnamefont {Dong-Ling}\
  \bibnamefont {Deng}}, \bibinfo {author} {\bibfnamefont {Xiaopeng}\
  \bibnamefont {Li}}, \ and\ \bibinfo {author} {\bibfnamefont {S.}~\bibnamefont
  {Das~Sarma}},\ }\bibfield  {title} {\enquote {\bibinfo {title} {Quantum
  entanglement in neural network states},}\ }\href {\doibase
  10.1103/PhysRevX.7.021021} {\bibfield  {journal} {\bibinfo  {journal} {Phys.
  Rev. X}\ }\textbf {\bibinfo {volume} {7}},\ \bibinfo {pages} {021021}
  (\bibinfo {year} {2017})}\BibitemShut {NoStop}%
\bibitem [{\citenamefont {Sharir}\ \emph {et~al.}(2022)\citenamefont {Sharir},
  \citenamefont {Shashua},\ and\ \citenamefont {Carleo}}]{Sharir_2022}%
  \BibitemOpen
  \bibfield  {author} {\bibinfo {author} {\bibfnamefont {Or}~\bibnamefont
  {Sharir}}, \bibinfo {author} {\bibfnamefont {Amnon}\ \bibnamefont {Shashua}},
  \ and\ \bibinfo {author} {\bibfnamefont {Giuseppe}\ \bibnamefont {Carleo}},\
  }\bibfield  {title} {\enquote {\bibinfo {title} {Neural tensor contractions
  and the expressive power of deep neural quantum states},}\ }\href {\doibase
  10.1103/PhysRevB.106.205136} {\bibfield  {journal} {\bibinfo  {journal}
  {Phys. Rev. B}\ }\textbf {\bibinfo {volume} {106}},\ \bibinfo {pages}
  {205136} (\bibinfo {year} {2022})}\BibitemShut {NoStop}%
\bibitem [{\citenamefont {Passetti}\ \emph {et~al.}(2023)\citenamefont
  {Passetti}, \citenamefont {Hofmann}, \citenamefont {Neitemeier},
  \citenamefont {Grunwald}, \citenamefont {Sentef},\ and\ \citenamefont
  {Kennes}}]{Passetti2023}%
  \BibitemOpen
  \bibfield  {author} {\bibinfo {author} {\bibfnamefont {Giacomo}\ \bibnamefont
  {Passetti}}, \bibinfo {author} {\bibfnamefont {Damian}\ \bibnamefont
  {Hofmann}}, \bibinfo {author} {\bibfnamefont {Pit}\ \bibnamefont
  {Neitemeier}}, \bibinfo {author} {\bibfnamefont {Lukas}\ \bibnamefont
  {Grunwald}}, \bibinfo {author} {\bibfnamefont {Michael~A.}\ \bibnamefont
  {Sentef}}, \ and\ \bibinfo {author} {\bibfnamefont {Dante~M.}\ \bibnamefont
  {Kennes}},\ }\bibfield  {title} {\enquote {\bibinfo {title} {Can neural
  quantum states learn volume-law ground states?}}\ }\href {\doibase
  10.1103/physrevlett.131.036502} {\bibfield  {journal} {\bibinfo  {journal}
  {Physical Review Letters}\ }\textbf {\bibinfo {volume} {131}},\ \bibinfo
  {pages} {036502} (\bibinfo {year} {2023})}\BibitemShut {NoStop}%
\bibitem [{\citenamefont {Halimeh}\ \emph
  {et~al.}(2025{\natexlab{a}})\citenamefont {Halimeh}, \citenamefont
  {Aidelsburger}, \citenamefont {Grusdt}, \citenamefont {Hauke},\ and\
  \citenamefont {Yang}}]{Halimeh2025}%
  \BibitemOpen
  \bibfield  {author} {\bibinfo {author} {\bibfnamefont {Jad~C.}\ \bibnamefont
  {Halimeh}}, \bibinfo {author} {\bibfnamefont {Monika}\ \bibnamefont
  {Aidelsburger}}, \bibinfo {author} {\bibfnamefont {Fabian}\ \bibnamefont
  {Grusdt}}, \bibinfo {author} {\bibfnamefont {Philipp}\ \bibnamefont {Hauke}},
  \ and\ \bibinfo {author} {\bibfnamefont {Bing}\ \bibnamefont {Yang}},\
  }\bibfield  {title} {\enquote {\bibinfo {title} {Cold-atom quantum simulators
  of gauge theories},}\ }\href {\doibase 10.1038/s41567-024-02721-8} {\bibfield
   {journal} {\bibinfo  {journal} {Nature Physics}\ }\textbf {\bibinfo {volume}
  {21}},\ \bibinfo {pages} {25--36} (\bibinfo {year}
  {2025}{\natexlab{a}})}\BibitemShut {NoStop}%
\bibitem [{\citenamefont {Zohar}\ \emph {et~al.}(2015)\citenamefont {Zohar},
  \citenamefont {Cirac},\ and\ \citenamefont {Reznik}}]{Zohar2015}%
  \BibitemOpen
  \bibfield  {author} {\bibinfo {author} {\bibfnamefont {Erez}\ \bibnamefont
  {Zohar}}, \bibinfo {author} {\bibfnamefont {J~Ignacio}\ \bibnamefont
  {Cirac}}, \ and\ \bibinfo {author} {\bibfnamefont {Benni}\ \bibnamefont
  {Reznik}},\ }\bibfield  {title} {\enquote {\bibinfo {title} {Quantum
  simulations of lattice gauge theories using ultracold atoms in optical
  lattices},}\ }\href {\doibase 10.1088/0034-4885/79/1/014401} {\bibfield
  {journal} {\bibinfo  {journal} {Reports on Progress in Physics}\ }\textbf
  {\bibinfo {volume} {79}},\ \bibinfo {pages} {014401} (\bibinfo {year}
  {2015})}\BibitemShut {NoStop}%
\bibitem [{\citenamefont {Wiese}(2013)}]{Wiese2013}%
  \BibitemOpen
  \bibfield  {author} {\bibinfo {author} {\bibfnamefont {U.‐J.}\ \bibnamefont
  {Wiese}},\ }\bibfield  {title} {\enquote {\bibinfo {title} {Ultracold quantum
  gases and lattice systems: quantum simulation of lattice gauge theories},}\
  }\href {\doibase 10.1002/andp.201300104} {\bibfield  {journal} {\bibinfo
  {journal} {Annalen der Physik}\ }\textbf {\bibinfo {volume} {525}},\ \bibinfo
  {pages} {777--796} (\bibinfo {year} {2013})}\BibitemShut {NoStop}%
\bibitem [{\citenamefont {Bauer}\ \emph
  {et~al.}(2023{\natexlab{a}})\citenamefont {Bauer}, \citenamefont {Davoudi},
  \citenamefont {Klco},\ and\ \citenamefont {Savage}}]{Bauer2023}%
  \BibitemOpen
  \bibfield  {author} {\bibinfo {author} {\bibfnamefont {Christian~W.}\
  \bibnamefont {Bauer}}, \bibinfo {author} {\bibfnamefont {Zohreh}\
  \bibnamefont {Davoudi}}, \bibinfo {author} {\bibfnamefont {Natalie}\
  \bibnamefont {Klco}}, \ and\ \bibinfo {author} {\bibfnamefont {Martin~J.}\
  \bibnamefont {Savage}},\ }\bibfield  {title} {\enquote {\bibinfo {title}
  {Quantum simulation of fundamental particles and forces},}\ }\href {\doibase
  10.1038/s42254-023-00599-8} {\bibfield  {journal} {\bibinfo  {journal}
  {Nature Reviews Physics}\ }\textbf {\bibinfo {volume} {5}},\ \bibinfo {pages}
  {420--432} (\bibinfo {year} {2023}{\natexlab{a}})}\BibitemShut {NoStop}%
\bibitem [{\citenamefont {Bauer}\ \emph
  {et~al.}(2023{\natexlab{b}})\citenamefont {Bauer}, \citenamefont {Davoudi},
  \citenamefont {Balantekin}, \citenamefont {Bhattacharya}, \citenamefont
  {Carena}, \citenamefont {de~Jong}, \citenamefont {Draper}, \citenamefont
  {El-Khadra}, \citenamefont {Gemelke}, \citenamefont {Hanada} \emph
  {et~al.}}]{Bauer2023qSimHEP}%
  \BibitemOpen
  \bibfield  {author} {\bibinfo {author} {\bibfnamefont {Christian~W.}\
  \bibnamefont {Bauer}}, \bibinfo {author} {\bibfnamefont {Zohreh}\
  \bibnamefont {Davoudi}}, \bibinfo {author} {\bibfnamefont {A.~Baha}\
  \bibnamefont {Balantekin}}, \bibinfo {author} {\bibfnamefont {Tanmoy}\
  \bibnamefont {Bhattacharya}}, \bibinfo {author} {\bibfnamefont {Marcela}\
  \bibnamefont {Carena}}, \bibinfo {author} {\bibfnamefont {Wibe~A.}\
  \bibnamefont {de~Jong}}, \bibinfo {author} {\bibfnamefont {Patrick}\
  \bibnamefont {Draper}}, \bibinfo {author} {\bibfnamefont {Aida}\ \bibnamefont
  {El-Khadra}}, \bibinfo {author} {\bibfnamefont {Nate}\ \bibnamefont
  {Gemelke}}, \bibinfo {author} {\bibfnamefont {Masanori}\ \bibnamefont
  {Hanada}},  \emph {et~al.},\ }\bibfield  {title} {\enquote {\bibinfo {title}
  {Quantum simulation for high-energy physics},}\ }\href {\doibase
  10.1103/prxquantum.4.027001} {\bibfield  {journal} {\bibinfo  {journal} {PRX
  Quantum}\ }\textbf {\bibinfo {volume} {4}},\ \bibinfo {pages} {027001}
  (\bibinfo {year} {2023}{\natexlab{b}})}\BibitemShut {NoStop}%
\bibitem [{\citenamefont {Di~Meglio}\ \emph {et~al.}(2024)\citenamefont
  {Di~Meglio}, \citenamefont {Jansen}, \citenamefont {Tavernelli},
  \citenamefont {Alexandrou}, \citenamefont {Arunachalam}, \citenamefont
  {Bauer}, \citenamefont {Borras}, \citenamefont {Carrazza}, \citenamefont
  {Crippa}, \citenamefont {Croft} \emph {et~al.}}]{DiMeglio2024}%
  \BibitemOpen
  \bibfield  {author} {\bibinfo {author} {\bibfnamefont {Alberto}\ \bibnamefont
  {Di~Meglio}}, \bibinfo {author} {\bibfnamefont {Karl}\ \bibnamefont
  {Jansen}}, \bibinfo {author} {\bibfnamefont {Ivano}\ \bibnamefont
  {Tavernelli}}, \bibinfo {author} {\bibfnamefont {Constantia}\ \bibnamefont
  {Alexandrou}}, \bibinfo {author} {\bibfnamefont {Srinivasan}\ \bibnamefont
  {Arunachalam}}, \bibinfo {author} {\bibfnamefont {Christian~W.}\ \bibnamefont
  {Bauer}}, \bibinfo {author} {\bibfnamefont {Kerstin}\ \bibnamefont {Borras}},
  \bibinfo {author} {\bibfnamefont {Stefano}\ \bibnamefont {Carrazza}},
  \bibinfo {author} {\bibfnamefont {Arianna}\ \bibnamefont {Crippa}}, \bibinfo
  {author} {\bibfnamefont {Vincent}\ \bibnamefont {Croft}},  \emph {et~al.},\
  }\bibfield  {title} {\enquote {\bibinfo {title} {Quantum computing for
  high-energy physics: State of the art and challenges},}\ }\href {\doibase
  10.1103/prxquantum.5.037001} {\bibfield  {journal} {\bibinfo  {journal} {PRX
  Quantum}\ }\textbf {\bibinfo {volume} {5}},\ \bibinfo {pages} {037001}
  (\bibinfo {year} {2024})}\BibitemShut {NoStop}%
\bibitem [{\citenamefont {Halimeh}\ \emph
  {et~al.}(2025{\natexlab{b}})\citenamefont {Halimeh}, \citenamefont {Mueller},
  \citenamefont {Knolle}, \citenamefont {Papić},\ and\ \citenamefont
  {Davoudi}}]{Halimeh2025QsimOutEquil}%
  \BibitemOpen
  \bibfield  {author} {\bibinfo {author} {\bibfnamefont {Jad~C.}\ \bibnamefont
  {Halimeh}}, \bibinfo {author} {\bibfnamefont {Niklas}\ \bibnamefont
  {Mueller}}, \bibinfo {author} {\bibfnamefont {Johannes}\ \bibnamefont
  {Knolle}}, \bibinfo {author} {\bibfnamefont {Zlatko}\ \bibnamefont {Papić}},
  \ and\ \bibinfo {author} {\bibfnamefont {Zohreh}\ \bibnamefont {Davoudi}},\
  }\href {\doibase 10.48550/ARXIV.2509.03586} {\enquote {\bibinfo {title}
  {Quantum simulation of out-of-equilibrium dynamics in gauge theories},}\ }
  (\bibinfo {year} {2025}{\natexlab{b}}),\ \Eprint
  {http://arxiv.org/abs/2509.03586} {arXiv:2509.03586 [quant-ph]} \BibitemShut
  {NoStop}%
\bibitem [{\citenamefont {Barbiero}\ \emph {et~al.}(2019)\citenamefont
  {Barbiero}, \citenamefont {Schweizer}, \citenamefont {Aidelsburger},
  \citenamefont {Demler}, \citenamefont {Goldman},\ and\ \citenamefont
  {Grusdt}}]{Barbiero2019}%
  \BibitemOpen
  \bibfield  {author} {\bibinfo {author} {\bibfnamefont {Luca}\ \bibnamefont
  {Barbiero}}, \bibinfo {author} {\bibfnamefont {Christian}\ \bibnamefont
  {Schweizer}}, \bibinfo {author} {\bibfnamefont {Monika}\ \bibnamefont
  {Aidelsburger}}, \bibinfo {author} {\bibfnamefont {Eugene}\ \bibnamefont
  {Demler}}, \bibinfo {author} {\bibfnamefont {Nathan}\ \bibnamefont
  {Goldman}}, \ and\ \bibinfo {author} {\bibfnamefont {Fabian}\ \bibnamefont
  {Grusdt}},\ }\bibfield  {title} {\enquote {\bibinfo {title} {Coupling
  ultracold matter to dynamical gauge fields in optical lattices: From flux
  attachment to $\mathbb{Z}_2$ lattice gauge theories},}\ }\href {\doibase
  10.1126/sciadv.aav7444} {\bibfield  {journal} {\bibinfo  {journal} {Science
  Advances}\ }\textbf {\bibinfo {volume} {5}} (\bibinfo {year} {2019}),\
  10.1126/sciadv.aav7444}\BibitemShut {NoStop}%
\bibitem [{\citenamefont {Görg}\ \emph {et~al.}(2019)\citenamefont {Görg},
  \citenamefont {Sandholzer}, \citenamefont {Minguzzi}, \citenamefont
  {Desbuquois}, \citenamefont {Messer},\ and\ \citenamefont
  {Esslinger}}]{Goerg2019}%
  \BibitemOpen
  \bibfield  {author} {\bibinfo {author} {\bibfnamefont {Frederik}\
  \bibnamefont {Görg}}, \bibinfo {author} {\bibfnamefont {Kilian}\
  \bibnamefont {Sandholzer}}, \bibinfo {author} {\bibfnamefont {Joaquín}\
  \bibnamefont {Minguzzi}}, \bibinfo {author} {\bibfnamefont {Rémi}\
  \bibnamefont {Desbuquois}}, \bibinfo {author} {\bibfnamefont {Michael}\
  \bibnamefont {Messer}}, \ and\ \bibinfo {author} {\bibfnamefont {Tilman}\
  \bibnamefont {Esslinger}},\ }\bibfield  {title} {\enquote {\bibinfo {title}
  {Realization of density-dependent {P}eierls phases to engineer quantized
  gauge fields coupled to ultracold matter},}\ }\href {\doibase
  10.1038/s41567-019-0615-4} {\bibfield  {journal} {\bibinfo  {journal} {Nature
  Physics}\ }\textbf {\bibinfo {volume} {15}},\ \bibinfo {pages} {1161--1167}
  (\bibinfo {year} {2019})}\BibitemShut {NoStop}%
\bibitem [{\citenamefont {Schweizer}\ \emph {et~al.}(2019)\citenamefont
  {Schweizer}, \citenamefont {Grusdt}, \citenamefont {Berngruber},
  \citenamefont {Barbiero}, \citenamefont {Demler}, \citenamefont {Goldman},
  \citenamefont {Bloch},\ and\ \citenamefont {Aidelsburger}}]{Schweizer2019}%
  \BibitemOpen
  \bibfield  {author} {\bibinfo {author} {\bibfnamefont {Christian}\
  \bibnamefont {Schweizer}}, \bibinfo {author} {\bibfnamefont {Fabian}\
  \bibnamefont {Grusdt}}, \bibinfo {author} {\bibfnamefont {Moritz}\
  \bibnamefont {Berngruber}}, \bibinfo {author} {\bibfnamefont {Luca}\
  \bibnamefont {Barbiero}}, \bibinfo {author} {\bibfnamefont {Eugene}\
  \bibnamefont {Demler}}, \bibinfo {author} {\bibfnamefont {Nathan}\
  \bibnamefont {Goldman}}, \bibinfo {author} {\bibfnamefont {Immanuel}\
  \bibnamefont {Bloch}}, \ and\ \bibinfo {author} {\bibfnamefont {Monika}\
  \bibnamefont {Aidelsburger}},\ }\bibfield  {title} {\enquote {\bibinfo
  {title} {Floquet approach to $\mathbb{Z}_2$ lattice gauge theories with
  ultracold atoms in optical lattices},}\ }\href {\doibase
  10.1038/s41567-019-0649-7} {\bibfield  {journal} {\bibinfo  {journal} {Nature
  Physics}\ }\textbf {\bibinfo {volume} {15}},\ \bibinfo {pages} {1168--1173}
  (\bibinfo {year} {2019})}\BibitemShut {NoStop}%
\bibitem [{\citenamefont {Mil}\ \emph {et~al.}(2020)\citenamefont {Mil},
  \citenamefont {Zache}, \citenamefont {Hegde}, \citenamefont {Xia},
  \citenamefont {Bhatt}, \citenamefont {Oberthaler}, \citenamefont {Hauke},
  \citenamefont {Berges},\ and\ \citenamefont {Jendrzejewski}}]{Mil2020}%
  \BibitemOpen
  \bibfield  {author} {\bibinfo {author} {\bibfnamefont {Alexander}\
  \bibnamefont {Mil}}, \bibinfo {author} {\bibfnamefont {Torsten~V.}\
  \bibnamefont {Zache}}, \bibinfo {author} {\bibfnamefont {Apoorva}\
  \bibnamefont {Hegde}}, \bibinfo {author} {\bibfnamefont {Andy}\ \bibnamefont
  {Xia}}, \bibinfo {author} {\bibfnamefont {Rohit~P.}\ \bibnamefont {Bhatt}},
  \bibinfo {author} {\bibfnamefont {Markus~K.}\ \bibnamefont {Oberthaler}},
  \bibinfo {author} {\bibfnamefont {Philipp}\ \bibnamefont {Hauke}}, \bibinfo
  {author} {\bibfnamefont {Jürgen}\ \bibnamefont {Berges}}, \ and\ \bibinfo
  {author} {\bibfnamefont {Fred}\ \bibnamefont {Jendrzejewski}},\ }\bibfield
  {title} {\enquote {\bibinfo {title} {A scalable realization of local {U(1)}
  gauge invariance in cold atomic mixtures},}\ }\href {\doibase
  10.1126/science.aaz5312} {\bibfield  {journal} {\bibinfo  {journal}
  {Science}\ }\textbf {\bibinfo {volume} {367}},\ \bibinfo {pages} {1128--1130}
  (\bibinfo {year} {2020})}\BibitemShut {NoStop}%
\bibitem [{\citenamefont {Yang}\ \emph {et~al.}(2020)\citenamefont {Yang},
  \citenamefont {Sun}, \citenamefont {Ott}, \citenamefont {Wang}, \citenamefont
  {Zache}, \citenamefont {Halimeh}, \citenamefont {Yuan}, \citenamefont
  {Hauke},\ and\ \citenamefont {Pan}}]{Yang2020}%
  \BibitemOpen
  \bibfield  {author} {\bibinfo {author} {\bibfnamefont {Bing}\ \bibnamefont
  {Yang}}, \bibinfo {author} {\bibfnamefont {Hui}\ \bibnamefont {Sun}},
  \bibinfo {author} {\bibfnamefont {Robert}\ \bibnamefont {Ott}}, \bibinfo
  {author} {\bibfnamefont {Han-Yi}\ \bibnamefont {Wang}}, \bibinfo {author}
  {\bibfnamefont {Torsten~V.}\ \bibnamefont {Zache}}, \bibinfo {author}
  {\bibfnamefont {Jad~C.}\ \bibnamefont {Halimeh}}, \bibinfo {author}
  {\bibfnamefont {Zhen-Sheng}\ \bibnamefont {Yuan}}, \bibinfo {author}
  {\bibfnamefont {Philipp}\ \bibnamefont {Hauke}}, \ and\ \bibinfo {author}
  {\bibfnamefont {Jian-Wei}\ \bibnamefont {Pan}},\ }\bibfield  {title}
  {\enquote {\bibinfo {title} {Observation of gauge invariance in a 71-site
  {B}ose–{H}ubbard quantum simulator},}\ }\href {\doibase
  10.1038/s41586-020-2910-8} {\bibfield  {journal} {\bibinfo  {journal}
  {Nature}\ }\textbf {\bibinfo {volume} {587}},\ \bibinfo {pages} {392--396}
  (\bibinfo {year} {2020})}\BibitemShut {NoStop}%
\bibitem [{\citenamefont {Zhou}\ \emph {et~al.}(2022)\citenamefont {Zhou},
  \citenamefont {Su}, \citenamefont {Halimeh}, \citenamefont {Ott},
  \citenamefont {Sun}, \citenamefont {Hauke}, \citenamefont {Yang},
  \citenamefont {Yuan}, \citenamefont {Berges},\ and\ \citenamefont
  {Pan}}]{Zhou2022}%
  \BibitemOpen
  \bibfield  {author} {\bibinfo {author} {\bibfnamefont {Zhao-Yu}\ \bibnamefont
  {Zhou}}, \bibinfo {author} {\bibfnamefont {Guo-Xian}\ \bibnamefont {Su}},
  \bibinfo {author} {\bibfnamefont {Jad~C.}\ \bibnamefont {Halimeh}}, \bibinfo
  {author} {\bibfnamefont {Robert}\ \bibnamefont {Ott}}, \bibinfo {author}
  {\bibfnamefont {Hui}\ \bibnamefont {Sun}}, \bibinfo {author} {\bibfnamefont
  {Philipp}\ \bibnamefont {Hauke}}, \bibinfo {author} {\bibfnamefont {Bing}\
  \bibnamefont {Yang}}, \bibinfo {author} {\bibfnamefont {Zhen-Sheng}\
  \bibnamefont {Yuan}}, \bibinfo {author} {\bibfnamefont {Jürgen}\
  \bibnamefont {Berges}}, \ and\ \bibinfo {author} {\bibfnamefont {Jian-Wei}\
  \bibnamefont {Pan}},\ }\bibfield  {title} {\enquote {\bibinfo {title}
  {Thermalization dynamics of a gauge theory on a quantum simulator},}\ }\href
  {\doibase 10.1126/science.abl6277} {\bibfield  {journal} {\bibinfo  {journal}
  {Science}\ }\textbf {\bibinfo {volume} {377}},\ \bibinfo {pages} {311--314}
  (\bibinfo {year} {2022})}\BibitemShut {NoStop}%
\bibitem [{\citenamefont {Kebrič}\ \emph {et~al.}(2025)\citenamefont
  {Kebrič}, \citenamefont {Su}, \citenamefont {Douglas}, \citenamefont
  {Szurek}, \citenamefont {Marković}, \citenamefont {Schollwöck},
  \citenamefont {Bohrdt}, \citenamefont {Greiner},\ and\ \citenamefont
  {Grusdt}}]{Kebric2025}%
  \BibitemOpen
  \bibfield  {author} {\bibinfo {author} {\bibfnamefont {Matjaž}\ \bibnamefont
  {Kebrič}}, \bibinfo {author} {\bibfnamefont {Lin}\ \bibnamefont {Su}},
  \bibinfo {author} {\bibfnamefont {Alexander}\ \bibnamefont {Douglas}},
  \bibinfo {author} {\bibfnamefont {Michal}\ \bibnamefont {Szurek}}, \bibinfo
  {author} {\bibfnamefont {Ognjen}\ \bibnamefont {Marković}}, \bibinfo
  {author} {\bibfnamefont {Ulrich}\ \bibnamefont {Schollwöck}}, \bibinfo
  {author} {\bibfnamefont {Annabelle}\ \bibnamefont {Bohrdt}}, \bibinfo
  {author} {\bibfnamefont {Markus}\ \bibnamefont {Greiner}}, \ and\ \bibinfo
  {author} {\bibfnamefont {Fabian}\ \bibnamefont {Grusdt}},\ }\href {\doibase
  10.48550/ARXIV.2509.16200} {\enquote {\bibinfo {title} {Exploring confinement
  transitions in $\mathbb{Z}_2$ lattice gauge theories with dipolar atoms
  beyond one dimension},}\ } (\bibinfo {year} {2025}),\ \Eprint
  {http://arxiv.org/abs/2509.16200} {arXiv:2509.16200 [cond-mat.quant-gas]}
  \BibitemShut {NoStop}%
\bibitem [{\citenamefont {Homeier}\ \emph {et~al.}(2023)\citenamefont
  {Homeier}, \citenamefont {Bohrdt}, \citenamefont {Linsel}, \citenamefont
  {Demler}, \citenamefont {Halimeh},\ and\ \citenamefont
  {Grusdt}}]{Homeier2023}%
  \BibitemOpen
  \bibfield  {author} {\bibinfo {author} {\bibfnamefont {Lukas}\ \bibnamefont
  {Homeier}}, \bibinfo {author} {\bibfnamefont {Annabelle}\ \bibnamefont
  {Bohrdt}}, \bibinfo {author} {\bibfnamefont {Simon}\ \bibnamefont {Linsel}},
  \bibinfo {author} {\bibfnamefont {Eugene}\ \bibnamefont {Demler}}, \bibinfo
  {author} {\bibfnamefont {Jad~C.}\ \bibnamefont {Halimeh}}, \ and\ \bibinfo
  {author} {\bibfnamefont {Fabian}\ \bibnamefont {Grusdt}},\ }\bibfield
  {title} {\enquote {\bibinfo {title} {Realistic scheme for quantum simulation
  of $\mathbb{Z}_{2}$ lattice gauge theories with dynamical matter in
  (2+1){D}},}\ }\href {\doibase 10.1038/s42005-023-01237-6} {\bibfield
  {journal} {\bibinfo  {journal} {Communications Physics}\ }\textbf {\bibinfo
  {volume} {6}} (\bibinfo {year} {2023}),\
  10.1038/s42005-023-01237-6}\BibitemShut {NoStop}%
\bibitem [{\citenamefont {Paredes}\ and\ \citenamefont
  {Bloch}(2008)}]{Paredes2008}%
  \BibitemOpen
  \bibfield  {author} {\bibinfo {author} {\bibfnamefont {Belén}\ \bibnamefont
  {Paredes}}\ and\ \bibinfo {author} {\bibfnamefont {Immanuel}\ \bibnamefont
  {Bloch}},\ }\bibfield  {title} {\enquote {\bibinfo {title} {Minimum instances
  of topological matter in an optical plaquette},}\ }\href {\doibase
  10.1103/physreva.77.023603} {\bibfield  {journal} {\bibinfo  {journal}
  {Physical Review A}\ }\textbf {\bibinfo {volume} {77}},\ \bibinfo {pages}
  {023603} (\bibinfo {year} {2008})}\BibitemShut {NoStop}%
\bibitem [{\citenamefont {Dai}\ \emph {et~al.}(2017)\citenamefont {Dai},
  \citenamefont {Yang}, \citenamefont {Reingruber}, \citenamefont {Sun},
  \citenamefont {Xu}, \citenamefont {Chen}, \citenamefont {Yuan},\ and\
  \citenamefont {Pan}}]{Dai2017}%
  \BibitemOpen
  \bibfield  {author} {\bibinfo {author} {\bibfnamefont {Han-Ning}\
  \bibnamefont {Dai}}, \bibinfo {author} {\bibfnamefont {Bing}\ \bibnamefont
  {Yang}}, \bibinfo {author} {\bibfnamefont {Andreas}\ \bibnamefont
  {Reingruber}}, \bibinfo {author} {\bibfnamefont {Hui}\ \bibnamefont {Sun}},
  \bibinfo {author} {\bibfnamefont {Xiao-Fan}\ \bibnamefont {Xu}}, \bibinfo
  {author} {\bibfnamefont {Yu-Ao}\ \bibnamefont {Chen}}, \bibinfo {author}
  {\bibfnamefont {Zhen-Sheng}\ \bibnamefont {Yuan}}, \ and\ \bibinfo {author}
  {\bibfnamefont {Jian-Wei}\ \bibnamefont {Pan}},\ }\bibfield  {title}
  {\enquote {\bibinfo {title} {Four-body ring-exchange interactions and anyonic
  statistics within a minimal toric-code {H}amiltonian},}\ }\href {\doibase
  10.1038/nphys4243} {\bibfield  {journal} {\bibinfo  {journal} {Nature
  Physics}\ }\textbf {\bibinfo {volume} {13}},\ \bibinfo {pages} {1195--1200}
  (\bibinfo {year} {2017})}\BibitemShut {NoStop}%
\bibitem [{\citenamefont {Tian}\ \emph {et~al.}(2025)\citenamefont {Tian},
  \citenamefont {Srivatsa}, \citenamefont {Xu}, \citenamefont {Osborne},
  \citenamefont {Borla},\ and\ \citenamefont {Halimeh}}]{Tian2025}%
  \BibitemOpen
  \bibfield  {author} {\bibinfo {author} {\bibfnamefont {Yizhuo}\ \bibnamefont
  {Tian}}, \bibinfo {author} {\bibfnamefont {N.~S.}\ \bibnamefont {Srivatsa}},
  \bibinfo {author} {\bibfnamefont {Kaidi}\ \bibnamefont {Xu}}, \bibinfo
  {author} {\bibfnamefont {Jesse~J.}\ \bibnamefont {Osborne}}, \bibinfo
  {author} {\bibfnamefont {Umberto}\ \bibnamefont {Borla}}, \ and\ \bibinfo
  {author} {\bibfnamefont {Jad~C.}\ \bibnamefont {Halimeh}},\ }\href {\doibase
  10.48550/ARXIV.2508.05736} {\enquote {\bibinfo {title} {Role of plaquette
  term in genuine $2+1$d string dynamics on quantum simulators},}\ } (\bibinfo
  {year} {2025}),\ \Eprint {http://arxiv.org/abs/2508.05736} {arXiv:2508.05736
  [quant-ph]} \BibitemShut {NoStop}%
\bibitem [{\citenamefont {Savary}\ and\ \citenamefont
  {Balents}(2016)}]{Savary2016}%
  \BibitemOpen
  \bibfield  {author} {\bibinfo {author} {\bibfnamefont {Lucile}\ \bibnamefont
  {Savary}}\ and\ \bibinfo {author} {\bibfnamefont {Leon}\ \bibnamefont
  {Balents}},\ }\bibfield  {title} {\enquote {\bibinfo {title} {Quantum spin
  liquids: a review},}\ }\href {\doibase 10.1088/0034-4885/80/1/016502}
  {\bibfield  {journal} {\bibinfo  {journal} {Reports on Progress in Physics}\
  }\textbf {\bibinfo {volume} {80}},\ \bibinfo {pages} {016502} (\bibinfo
  {year} {2016})}\BibitemShut {NoStop}%
\bibitem [{\citenamefont {Martinez}\ \emph {et~al.}(2016)\citenamefont
  {Martinez}, \citenamefont {Muschik}, \citenamefont {Schindler}, \citenamefont
  {Nigg}, \citenamefont {Erhard}, \citenamefont {Heyl}, \citenamefont {Hauke},
  \citenamefont {Dalmonte}, \citenamefont {Monz}, \citenamefont {Zoller},\ and\
  \citenamefont {Blatt}}]{Martinez2016}%
  \BibitemOpen
  \bibfield  {author} {\bibinfo {author} {\bibfnamefont {Esteban~A.}\
  \bibnamefont {Martinez}}, \bibinfo {author} {\bibfnamefont {Christine~A.}\
  \bibnamefont {Muschik}}, \bibinfo {author} {\bibfnamefont {Philipp}\
  \bibnamefont {Schindler}}, \bibinfo {author} {\bibfnamefont {Daniel}\
  \bibnamefont {Nigg}}, \bibinfo {author} {\bibfnamefont {Alexander}\
  \bibnamefont {Erhard}}, \bibinfo {author} {\bibfnamefont {Markus}\
  \bibnamefont {Heyl}}, \bibinfo {author} {\bibfnamefont {Philipp}\
  \bibnamefont {Hauke}}, \bibinfo {author} {\bibfnamefont {Marcello}\
  \bibnamefont {Dalmonte}}, \bibinfo {author} {\bibfnamefont {Thomas}\
  \bibnamefont {Monz}}, \bibinfo {author} {\bibfnamefont {Peter}\ \bibnamefont
  {Zoller}}, \ and\ \bibinfo {author} {\bibfnamefont {Rainer}\ \bibnamefont
  {Blatt}},\ }\bibfield  {title} {\enquote {\bibinfo {title} {Real-time
  dynamics of lattice gauge theories with a few-qubit quantum computer},}\
  }\href {\doibase 10.1038/nature18318} {\bibfield  {journal} {\bibinfo
  {journal} {Nature}\ }\textbf {\bibinfo {volume} {534}},\ \bibinfo {pages}
  {516--519} (\bibinfo {year} {2016})}\BibitemShut {NoStop}%
\bibitem [{\citenamefont {Gyawali}\ \emph {et~al.}(2024)\citenamefont
  {Gyawali}, \citenamefont {Kumar}, \citenamefont {Lensky}, \citenamefont
  {Rosenberg}, \citenamefont {Szasz}, \citenamefont {Cochran}, \citenamefont
  {Chen}, \citenamefont {Karamlou}, \citenamefont {Kechedzhi}, \citenamefont
  {Berndtsson} \emph {et~al.}}]{Gyawali2024}%
  \BibitemOpen
  \bibfield  {author} {\bibinfo {author} {\bibfnamefont {Gaurav}\ \bibnamefont
  {Gyawali}}, \bibinfo {author} {\bibfnamefont {Shashwat}\ \bibnamefont
  {Kumar}}, \bibinfo {author} {\bibfnamefont {Yuri~D.}\ \bibnamefont {Lensky}},
  \bibinfo {author} {\bibfnamefont {Eliott}\ \bibnamefont {Rosenberg}},
  \bibinfo {author} {\bibfnamefont {Aaron}\ \bibnamefont {Szasz}}, \bibinfo
  {author} {\bibfnamefont {Tyler}\ \bibnamefont {Cochran}}, \bibinfo {author}
  {\bibfnamefont {Renyi}\ \bibnamefont {Chen}}, \bibinfo {author}
  {\bibfnamefont {Amir~H.}\ \bibnamefont {Karamlou}}, \bibinfo {author}
  {\bibfnamefont {Kostyantyn}\ \bibnamefont {Kechedzhi}}, \bibinfo {author}
  {\bibfnamefont {Julia}\ \bibnamefont {Berndtsson}},  \emph {et~al.},\ }\href
  {\doibase 10.48550/ARXIV.2410.06557} {\enquote {\bibinfo {title} {Observation
  of disorder-free localization using a (2+1){D} lattice gauge theory on a
  quantum processor},}\ } (\bibinfo {year} {2024}),\ \Eprint
  {http://arxiv.org/abs/2410.06557} {arXiv:2410.06557 [quant-ph]} \BibitemShut
  {NoStop}%
\bibitem [{\citenamefont {Cochran}\ \emph {et~al.}(2025)\citenamefont
  {Cochran}, \citenamefont {Jobst}, \citenamefont {Rosenberg}, \citenamefont
  {Lensky}, \citenamefont {Gyawali}, \citenamefont {Eassa}, \citenamefont
  {Will}, \citenamefont {Szasz}, \citenamefont {Abanin}, \citenamefont
  {Acharya} \emph {et~al.}}]{Cochran2025}%
  \BibitemOpen
  \bibfield  {author} {\bibinfo {author} {\bibfnamefont {T.~A.}\ \bibnamefont
  {Cochran}}, \bibinfo {author} {\bibfnamefont {B.}~\bibnamefont {Jobst}},
  \bibinfo {author} {\bibfnamefont {E.}~\bibnamefont {Rosenberg}}, \bibinfo
  {author} {\bibfnamefont {Y.~D.}\ \bibnamefont {Lensky}}, \bibinfo {author}
  {\bibfnamefont {G.}~\bibnamefont {Gyawali}}, \bibinfo {author} {\bibfnamefont
  {N.}~\bibnamefont {Eassa}}, \bibinfo {author} {\bibfnamefont
  {M.}~\bibnamefont {Will}}, \bibinfo {author} {\bibfnamefont {A.}~\bibnamefont
  {Szasz}}, \bibinfo {author} {\bibfnamefont {D.}~\bibnamefont {Abanin}},
  \bibinfo {author} {\bibfnamefont {R.}~\bibnamefont {Acharya}},  \emph
  {et~al.},\ }\bibfield  {title} {\enquote {\bibinfo {title} {Visualizing
  dynamics of charges and strings in {(2 + 1)D} lattice gauge theories},}\
  }\href {\doibase 10.1038/s41586-025-08999-9} {\bibfield  {journal} {\bibinfo
  {journal} {Nature}\ }\textbf {\bibinfo {volume} {642}},\ \bibinfo {pages}
  {315--320} (\bibinfo {year} {2025})}\BibitemShut {NoStop}%
\bibitem [{\citenamefont {Mildenberger}\ \emph {et~al.}(2025)\citenamefont
  {Mildenberger}, \citenamefont {Mruczkiewicz}, \citenamefont {Halimeh},
  \citenamefont {Jiang},\ and\ \citenamefont {Hauke}}]{Mildenberger2025}%
  \BibitemOpen
  \bibfield  {author} {\bibinfo {author} {\bibfnamefont {Julius}\ \bibnamefont
  {Mildenberger}}, \bibinfo {author} {\bibfnamefont {Wojciech}\ \bibnamefont
  {Mruczkiewicz}}, \bibinfo {author} {\bibfnamefont {Jad~C.}\ \bibnamefont
  {Halimeh}}, \bibinfo {author} {\bibfnamefont {Zhang}\ \bibnamefont {Jiang}},
  \ and\ \bibinfo {author} {\bibfnamefont {Philipp}\ \bibnamefont {Hauke}},\
  }\bibfield  {title} {\enquote {\bibinfo {title} {Confinement in a
  $\mathbb{Z}_{2}$ lattice gauge theory on a quantum computer},}\ }\href
  {\doibase 10.1038/s41567-024-02723-6} {\bibfield  {journal} {\bibinfo
  {journal} {Nature Physics}\ }\textbf {\bibinfo {volume} {21}},\ \bibinfo
  {pages} {312--317} (\bibinfo {year} {2025})}\BibitemShut {NoStop}%
\bibitem [{\citenamefont {González-Cuadra}\ \emph {et~al.}(2025)\citenamefont
  {González-Cuadra}, \citenamefont {Hamdan}, \citenamefont {Zache},
  \citenamefont {Braverman}, \citenamefont {Kornjača}, \citenamefont {Lukin},
  \citenamefont {Cantú}, \citenamefont {Liu}, \citenamefont {Wang},
  \citenamefont {Keesling}, \citenamefont {Lukin}, \citenamefont {Zoller},\
  and\ \citenamefont {Bylinskii}}]{GonzalezCuadra2025}%
  \BibitemOpen
  \bibfield  {author} {\bibinfo {author} {\bibfnamefont {Daniel}\ \bibnamefont
  {González-Cuadra}}, \bibinfo {author} {\bibfnamefont {Majd}\ \bibnamefont
  {Hamdan}}, \bibinfo {author} {\bibfnamefont {Torsten~V.}\ \bibnamefont
  {Zache}}, \bibinfo {author} {\bibfnamefont {Boris}\ \bibnamefont
  {Braverman}}, \bibinfo {author} {\bibfnamefont {Milan}\ \bibnamefont
  {Kornjača}}, \bibinfo {author} {\bibfnamefont {Alexander}\ \bibnamefont
  {Lukin}}, \bibinfo {author} {\bibfnamefont {Sergio~H.}\ \bibnamefont
  {Cantú}}, \bibinfo {author} {\bibfnamefont {Fangli}\ \bibnamefont {Liu}},
  \bibinfo {author} {\bibfnamefont {Sheng-Tao}\ \bibnamefont {Wang}}, \bibinfo
  {author} {\bibfnamefont {Alexander}\ \bibnamefont {Keesling}}, \bibinfo
  {author} {\bibfnamefont {Mikhail~D.}\ \bibnamefont {Lukin}}, \bibinfo
  {author} {\bibfnamefont {Peter}\ \bibnamefont {Zoller}}, \ and\ \bibinfo
  {author} {\bibfnamefont {Alexei}\ \bibnamefont {Bylinskii}},\ }\bibfield
  {title} {\enquote {\bibinfo {title} {Observation of string breaking on a {(2
  + 1)D} {R}ydberg quantum simulator},}\ }\href {\doibase
  10.1038/s41586-025-09051-6} {\bibfield  {journal} {\bibinfo  {journal}
  {Nature}\ }\textbf {\bibinfo {volume} {642}},\ \bibinfo {pages} {321--326}
  (\bibinfo {year} {2025})}\BibitemShut {NoStop}%
\bibitem [{\citenamefont {De}\ \emph {et~al.}(2024)\citenamefont {De},
  \citenamefont {Lerose}, \citenamefont {Luo}, \citenamefont {Surace},
  \citenamefont {Schuckert}, \citenamefont {Bennewitz}, \citenamefont {Ware},
  \citenamefont {Morong}, \citenamefont {Collins}, \citenamefont {Davoudi},
  \citenamefont {Gorshkov}, \citenamefont {Katz},\ and\ \citenamefont
  {Monroe}}]{De2024}%
  \BibitemOpen
  \bibfield  {author} {\bibinfo {author} {\bibfnamefont {Arinjoy}\ \bibnamefont
  {De}}, \bibinfo {author} {\bibfnamefont {Alessio}\ \bibnamefont {Lerose}},
  \bibinfo {author} {\bibfnamefont {De}~\bibnamefont {Luo}}, \bibinfo {author}
  {\bibfnamefont {Federica~M.}\ \bibnamefont {Surace}}, \bibinfo {author}
  {\bibfnamefont {Alexander}\ \bibnamefont {Schuckert}}, \bibinfo {author}
  {\bibfnamefont {Elizabeth~R.}\ \bibnamefont {Bennewitz}}, \bibinfo {author}
  {\bibfnamefont {Brayden}\ \bibnamefont {Ware}}, \bibinfo {author}
  {\bibfnamefont {William}\ \bibnamefont {Morong}}, \bibinfo {author}
  {\bibfnamefont {Kate~S.}\ \bibnamefont {Collins}}, \bibinfo {author}
  {\bibfnamefont {Zohreh}\ \bibnamefont {Davoudi}}, \bibinfo {author}
  {\bibfnamefont {Alexey~V.}\ \bibnamefont {Gorshkov}}, \bibinfo {author}
  {\bibfnamefont {Or}~\bibnamefont {Katz}}, \ and\ \bibinfo {author}
  {\bibfnamefont {Christopher}\ \bibnamefont {Monroe}},\ }\href {\doibase
  10.48550/ARXIV.2410.13815} {\enquote {\bibinfo {title} {Observation of
  string-breaking dynamics in a quantum simulator},}\ } (\bibinfo {year}
  {2024}),\ \Eprint {http://arxiv.org/abs/2410.13815} {arXiv:2410.13815
  [quant-ph]} \BibitemShut {NoStop}%
\bibitem [{\citenamefont {Hubig}\ \emph {et~al.}(2023)\citenamefont {Hubig},
  \citenamefont {Lachenmaier}, \citenamefont {Linden}, \citenamefont
  {Reinhard}, \citenamefont {Stenzel}, \citenamefont {Swoboda}, \citenamefont
  {Grundner},\ and\ \citenamefont {Mardazad}}]{hubig:_syten_toolk}%
  \BibitemOpen
  \bibfield  {author} {\bibinfo {author} {\bibfnamefont {Claudius}\
  \bibnamefont {Hubig}}, \bibinfo {author} {\bibfnamefont {Felix}\ \bibnamefont
  {Lachenmaier}}, \bibinfo {author} {\bibfnamefont {Nils-Oliver}\ \bibnamefont
  {Linden}}, \bibinfo {author} {\bibfnamefont {Teresa}\ \bibnamefont
  {Reinhard}}, \bibinfo {author} {\bibfnamefont {Leo}\ \bibnamefont {Stenzel}},
  \bibinfo {author} {\bibfnamefont {Andreas}\ \bibnamefont {Swoboda}}, \bibinfo
  {author} {\bibfnamefont {Martin}\ \bibnamefont {Grundner}}, \ and\ \bibinfo
  {author} {\bibfnamefont {Sam}\ \bibnamefont {Mardazad}},\ }\href
  {https://syten.eu} {\enquote {\bibinfo {title} {The \textsc{SyTen}
  toolkit},}\ } (\bibinfo {year} {2023})\BibitemShut {NoStop}%
\bibitem [{\citenamefont
  {Hubig}(2017)}]{hubig17:_symmet_protec_tensor_network}%
  \BibitemOpen
  \bibfield  {author} {\bibinfo {author} {\bibfnamefont {Claudius}\
  \bibnamefont {Hubig}},\ }\emph {\bibinfo {title} {Symmetry-Protected Tensor
  Networks}},\ \href {https://edoc.ub.uni-muenchen.de/21348/} {Ph.D. thesis},\
  \bibinfo  {school} {LMU München} (\bibinfo {year} {2017})\BibitemShut
  {NoStop}%
\bibitem [{\citenamefont {Wang}\ and\ \citenamefont {Pollet}(2025)}]{Wang2025}%
  \BibitemOpen
  \bibfield  {author} {\bibinfo {author} {\bibfnamefont {Zhenjiu}\ \bibnamefont
  {Wang}}\ and\ \bibinfo {author} {\bibfnamefont {Lode}\ \bibnamefont
  {Pollet}},\ }\bibfield  {title} {\enquote {\bibinfo {title} {Renormalized
  classical spin liquid on the {R}uby lattice},}\ }\href {\doibase
  10.1103/physrevlett.134.086601} {\bibfield  {journal} {\bibinfo  {journal}
  {Physical Review Letters}\ }\textbf {\bibinfo {volume} {134}},\ \bibinfo
  {pages} {086601} (\bibinfo {year} {2025})}\BibitemShut {NoStop}%
\bibitem [{\citenamefont {Prosko}\ \emph {et~al.}(2017)\citenamefont {Prosko},
  \citenamefont {Lee},\ and\ \citenamefont {Maciejko}}]{Prosko2017}%
  \BibitemOpen
  \bibfield  {author} {\bibinfo {author} {\bibfnamefont {Christian}\
  \bibnamefont {Prosko}}, \bibinfo {author} {\bibfnamefont {Shu-Ping}\
  \bibnamefont {Lee}}, \ and\ \bibinfo {author} {\bibfnamefont {Joseph}\
  \bibnamefont {Maciejko}},\ }\bibfield  {title} {\enquote {\bibinfo {title}
  {Simple $\mathbb{Z}_2$ lattice gauge theories at finite fermion density},}\
  }\href {\doibase 10.1103/physrevb.96.205104} {\bibfield  {journal} {\bibinfo
  {journal} {Physical Review B}\ }\textbf {\bibinfo {volume} {96}},\ \bibinfo
  {pages} {205104} (\bibinfo {year} {2017})}\BibitemShut {NoStop}%
\bibitem [{SM()}]{SM}%
  \BibitemOpen
  \href@noop {} {}\bibinfo {note} {See Supplemental Material at [\emph{URL will
  be inserted by publisher}] for details on the mapping of the (2+1){D}
  $\mathbb{Z}_2$ LGT to the spin model, details on numerical simulations, and
  details on the pair-pair correlations.}\BibitemShut {Stop}%
\bibitem [{\citenamefont {Borla}\ \emph {et~al.}(2020)\citenamefont {Borla},
  \citenamefont {Verresen}, \citenamefont {Grusdt},\ and\ \citenamefont
  {Moroz}}]{Borla2020}%
  \BibitemOpen
  \bibfield  {author} {\bibinfo {author} {\bibfnamefont {Umberto}\ \bibnamefont
  {Borla}}, \bibinfo {author} {\bibfnamefont {Ruben}\ \bibnamefont {Verresen}},
  \bibinfo {author} {\bibfnamefont {Fabian}\ \bibnamefont {Grusdt}}, \ and\
  \bibinfo {author} {\bibfnamefont {Sergej}\ \bibnamefont {Moroz}},\ }\bibfield
   {title} {\enquote {\bibinfo {title} {Confined phases of one-dimensional
  spinless fermions coupled to {$Z_2$} gauge theory},}\ }\href {\doibase
  10.1103/physrevlett.124.120503} {\bibfield  {journal} {\bibinfo  {journal}
  {Physical Review Letters}\ }\textbf {\bibinfo {volume} {124}},\ \bibinfo
  {pages} {120503} (\bibinfo {year} {2020})}\BibitemShut {NoStop}%
\bibitem [{\citenamefont {Kebrič}\ \emph {et~al.}(2021)\citenamefont
  {Kebrič}, \citenamefont {Barbiero}, \citenamefont {Reinmoser}, \citenamefont
  {Schollwöck},\ and\ \citenamefont {Grusdt}}]{Kebric2021}%
  \BibitemOpen
  \bibfield  {author} {\bibinfo {author} {\bibfnamefont {Matjaž}\ \bibnamefont
  {Kebrič}}, \bibinfo {author} {\bibfnamefont {Luca}\ \bibnamefont
  {Barbiero}}, \bibinfo {author} {\bibfnamefont {Christian}\ \bibnamefont
  {Reinmoser}}, \bibinfo {author} {\bibfnamefont {Ulrich}\ \bibnamefont
  {Schollwöck}}, \ and\ \bibinfo {author} {\bibfnamefont {Fabian}\
  \bibnamefont {Grusdt}},\ }\bibfield  {title} {\enquote {\bibinfo {title}
  {Confinement and mott transitions of dynamical charges in one-dimensional
  lattice gauge theories},}\ }\href {\doibase 10.1103/physrevlett.127.167203}
  {\bibfield  {journal} {\bibinfo  {journal} {Physical Review Letters}\
  }\textbf {\bibinfo {volume} {127}},\ \bibinfo {pages} {167203} (\bibinfo
  {year} {2021})}\BibitemShut {NoStop}%
\bibitem [{\citenamefont {Ding}(1992)}]{Ding1992}%
  \BibitemOpen
  \bibfield  {author} {\bibinfo {author} {\bibfnamefont {H.-Q.}\ \bibnamefont
  {Ding}},\ }\bibfield  {title} {\enquote {\bibinfo {title} {Phase transition
  and thermodynamics of quantum {XY} model in two dimensions},}\ }\href
  {\doibase 10.1103/physrevb.45.230} {\bibfield  {journal} {\bibinfo  {journal}
  {Physical Review B}\ }\textbf {\bibinfo {volume} {45}},\ \bibinfo {pages}
  {230--242} (\bibinfo {year} {1992})}\BibitemShut {NoStop}%
\bibitem [{\citenamefont {Borla}\ \emph {et~al.}(2022)\citenamefont {Borla},
  \citenamefont {Jeevanesan}, \citenamefont {Pollmann},\ and\ \citenamefont
  {Moroz}}]{Borla2022}%
  \BibitemOpen
  \bibfield  {author} {\bibinfo {author} {\bibfnamefont {Umberto}\ \bibnamefont
  {Borla}}, \bibinfo {author} {\bibfnamefont {Bhilahari}\ \bibnamefont
  {Jeevanesan}}, \bibinfo {author} {\bibfnamefont {Frank}\ \bibnamefont
  {Pollmann}}, \ and\ \bibinfo {author} {\bibfnamefont {Sergej}\ \bibnamefont
  {Moroz}},\ }\bibfield  {title} {\enquote {\bibinfo {title} {Quantum phases of
  two-dimensional $\mathbb{Z}_2$ gauge theory coupled to single-component
  fermion matter},}\ }\href {\doibase 10.1103/physrevb.105.075132} {\bibfield
  {journal} {\bibinfo  {journal} {Physical Review B}\ }\textbf {\bibinfo
  {volume} {105}},\ \bibinfo {pages} {075132} (\bibinfo {year}
  {2022})}\BibitemShut {NoStop}%
\bibitem [{\citenamefont {Borla}\ \emph {et~al.}(2024)\citenamefont {Borla},
  \citenamefont {Gazit},\ and\ \citenamefont {Moroz}}]{Borla2024}%
  \BibitemOpen
  \bibfield  {author} {\bibinfo {author} {\bibfnamefont {Umberto}\ \bibnamefont
  {Borla}}, \bibinfo {author} {\bibfnamefont {Snir}\ \bibnamefont {Gazit}}, \
  and\ \bibinfo {author} {\bibfnamefont {Sergej}\ \bibnamefont {Moroz}},\
  }\bibfield  {title} {\enquote {\bibinfo {title} {Deconfined quantum
  criticality in {I}sing gauge theory entangled with single-component
  fermions},}\ }\href {\doibase 10.1103/physrevb.110.l201110} {\bibfield
  {journal} {\bibinfo  {journal} {Physical Review B}\ }\textbf {\bibinfo
  {volume} {110}},\ \bibinfo {pages} {l201110} (\bibinfo {year}
  {2024})}\BibitemShut {NoStop}%
\bibitem [{\citenamefont {Kebrič}\ \emph {et~al.}(2023)\citenamefont
  {Kebrič}, \citenamefont {Borla}, \citenamefont {Schollwöck}, \citenamefont
  {Moroz}, \citenamefont {Barbiero},\ and\ \citenamefont
  {Grusdt}}]{Kebric2023}%
  \BibitemOpen
  \bibfield  {author} {\bibinfo {author} {\bibfnamefont {Matjaž}\ \bibnamefont
  {Kebrič}}, \bibinfo {author} {\bibfnamefont {Umberto}\ \bibnamefont
  {Borla}}, \bibinfo {author} {\bibfnamefont {Ulrich}\ \bibnamefont
  {Schollwöck}}, \bibinfo {author} {\bibfnamefont {Sergej}\ \bibnamefont
  {Moroz}}, \bibinfo {author} {\bibfnamefont {Luca}\ \bibnamefont {Barbiero}},
  \ and\ \bibinfo {author} {\bibfnamefont {Fabian}\ \bibnamefont {Grusdt}},\
  }\bibfield  {title} {\enquote {\bibinfo {title} {Confinement induced
  frustration in a one-dimensional $\mathbb{Z}_2$ lattice gauge theory},}\
  }\href {\doibase 10.1088/1367-2630/acb45c} {\bibfield  {journal} {\bibinfo
  {journal} {New Journal of Physics}\ }\textbf {\bibinfo {volume} {25}},\
  \bibinfo {pages} {013035} (\bibinfo {year} {2023})}\BibitemShut {NoStop}%
\bibitem [{\citenamefont {Fredenhagen}\ and\ \citenamefont
  {Marcu}(1983)}]{Fredenhagen1983}%
  \BibitemOpen
  \bibfield  {author} {\bibinfo {author} {\bibfnamefont {Klaus}\ \bibnamefont
  {Fredenhagen}}\ and\ \bibinfo {author} {\bibfnamefont {Mihail}\ \bibnamefont
  {Marcu}},\ }\bibfield  {title} {\enquote {\bibinfo {title} {Charged states in
  $\mathbb{Z}_2$ gauge theories},}\ }\href {\doibase 10.1007/bf01206315}
  {\bibfield  {journal} {\bibinfo  {journal} {Communications in Mathematical
  Physics}\ }\textbf {\bibinfo {volume} {92}},\ \bibinfo {pages} {81--119}
  (\bibinfo {year} {1983})}\BibitemShut {NoStop}%
\bibitem [{\citenamefont {Fredenhagen}\ and\ \citenamefont
  {Marcu}(1986)}]{Fredenhagen1986}%
  \BibitemOpen
  \bibfield  {author} {\bibinfo {author} {\bibfnamefont {Klaus}\ \bibnamefont
  {Fredenhagen}}\ and\ \bibinfo {author} {\bibfnamefont {Mihail}\ \bibnamefont
  {Marcu}},\ }\bibfield  {title} {\enquote {\bibinfo {title} {Confinement
  criterion for {QCD} with dynamical quarks},}\ }\href {\doibase
  10.1103/physrevlett.56.223} {\bibfield  {journal} {\bibinfo  {journal}
  {Physical Review Letters}\ }\textbf {\bibinfo {volume} {56}},\ \bibinfo
  {pages} {223--224} (\bibinfo {year} {1986})}\BibitemShut {NoStop}%
\bibitem [{\citenamefont {Marcu}(1986)}]{Marcu1986}%
  \BibitemOpen
  \bibfield  {author} {\bibinfo {author} {\bibfnamefont {Mihail}\ \bibnamefont
  {Marcu}},\ }\enquote {\bibinfo {title} {({U}ses of) {A}n order parameter for
  lattice gauge theories with matter fields},}\ in\ \href {\doibase
  10.1007/978-1-4613-2231-3_25} {\emph {\bibinfo {booktitle} {Lattice Gauge
  Theory}}}\ (\bibinfo  {publisher} {Springer US},\ \bibinfo {year} {1986})\
  pp.\ \bibinfo {pages} {267--278}\BibitemShut {NoStop}%
\bibitem [{\citenamefont {Fredenhagen}\ and\ \citenamefont
  {Marcu}(1988)}]{Fredenhagen1988}%
  \BibitemOpen
  \bibfield  {author} {\bibinfo {author} {\bibfnamefont {Klaus}\ \bibnamefont
  {Fredenhagen}}\ and\ \bibinfo {author} {\bibfnamefont {Mihail}\ \bibnamefont
  {Marcu}},\ }\bibfield  {title} {\enquote {\bibinfo {title} {Dual
  interpretation of order parameters for lattice gauge theories with matter
  fields},}\ }\href {\doibase 10.1016/0920-5632(88)90124-7} {\bibfield
  {journal} {\bibinfo  {journal} {Nuclear Physics B - Proceedings Supplements}\
  }\textbf {\bibinfo {volume} {4}},\ \bibinfo {pages} {352--357} (\bibinfo
  {year} {1988})}\BibitemShut {NoStop}%
\bibitem [{\citenamefont {Cobos}\ \emph {et~al.}(2025)\citenamefont {Cobos},
  \citenamefont {Fraxanet}, \citenamefont {Benito}, \citenamefont
  {di~Marcantonio}, \citenamefont {Rivero}, \citenamefont {Kapás},
  \citenamefont {Werner}, \citenamefont {Legeza}, \citenamefont {Bermudez},\
  and\ \citenamefont {Rico}}]{Cobos2025}%
  \BibitemOpen
  \bibfield  {author} {\bibinfo {author} {\bibfnamefont {Jesús}\ \bibnamefont
  {Cobos}}, \bibinfo {author} {\bibfnamefont {Joana}\ \bibnamefont {Fraxanet}},
  \bibinfo {author} {\bibfnamefont {César}\ \bibnamefont {Benito}}, \bibinfo
  {author} {\bibfnamefont {Francesco}\ \bibnamefont {di~Marcantonio}}, \bibinfo
  {author} {\bibfnamefont {Pedro}\ \bibnamefont {Rivero}}, \bibinfo {author}
  {\bibfnamefont {Kornél}\ \bibnamefont {Kapás}}, \bibinfo {author}
  {\bibfnamefont {Miklós~Antal}\ \bibnamefont {Werner}}, \bibinfo {author}
  {\bibfnamefont {Örs}\ \bibnamefont {Legeza}}, \bibinfo {author}
  {\bibfnamefont {Alejandro}\ \bibnamefont {Bermudez}}, \ and\ \bibinfo
  {author} {\bibfnamefont {Enrique}\ \bibnamefont {Rico}},\ }\href {\doibase
  10.48550/ARXIV.2507.08088} {\enquote {\bibinfo {title} {Real-time dynamics in
  a (2+1)-{D} gauge theory: The stringy nature on a superconducting quantum
  simulator},}\ } (\bibinfo {year} {2025}),\ \Eprint
  {http://arxiv.org/abs/2507.08088} {arXiv:2507.08088 [quant-ph]} \BibitemShut
  {NoStop}%
\bibitem [{\citenamefont {Homeier}\ \emph {et~al.}(2021)\citenamefont
  {Homeier}, \citenamefont {Schweizer}, \citenamefont {Aidelsburger},
  \citenamefont {Fedorov},\ and\ \citenamefont {Grusdt}}]{Homeier2021}%
  \BibitemOpen
  \bibfield  {author} {\bibinfo {author} {\bibfnamefont {Lukas}\ \bibnamefont
  {Homeier}}, \bibinfo {author} {\bibfnamefont {Christian}\ \bibnamefont
  {Schweizer}}, \bibinfo {author} {\bibfnamefont {Monika}\ \bibnamefont
  {Aidelsburger}}, \bibinfo {author} {\bibfnamefont {Arkady}\ \bibnamefont
  {Fedorov}}, \ and\ \bibinfo {author} {\bibfnamefont {Fabian}\ \bibnamefont
  {Grusdt}},\ }\bibfield  {title} {\enquote {\bibinfo {title} {$\mathbb{Z}_2$
  lattice gauge theories and {K}itaev’s toric code: A scheme for analog
  quantum simulation},}\ }\href {\doibase 10.1103/physrevb.104.085138}
  {\bibfield  {journal} {\bibinfo  {journal} {Physical Review B}\ }\textbf
  {\bibinfo {volume} {104}},\ \bibinfo {pages} {085138} (\bibinfo {year}
  {2021})}\BibitemShut {NoStop}%
\bibitem [{\citenamefont {Döschl}\ and\ \citenamefont
  {Bohrdt}(2025)}]{Doeschl_2025_Correlations}%
  \BibitemOpen
  \bibfield  {author} {\bibinfo {author} {\bibfnamefont {Fabian}\ \bibnamefont
  {Döschl}}\ and\ \bibinfo {author} {\bibfnamefont {Annabelle}\ \bibnamefont
  {Bohrdt}},\ }\href {https://arxiv.org/abs/2508.14152} {\enquote {\bibinfo
  {title} {Importance of correlations for neural quantum states},}\ } (\bibinfo
  {year} {2025}),\ \Eprint {http://arxiv.org/abs/2508.14152} {arXiv:2508.14152
  [quant-ph]} \BibitemShut {NoStop}%
\bibitem [{\citenamefont {Luo}\ \emph {et~al.}(2021)\citenamefont {Luo},
  \citenamefont {Carleo}, \citenamefont {Clark},\ and\ \citenamefont
  {Stokes}}]{Luo2021}%
  \BibitemOpen
  \bibfield  {author} {\bibinfo {author} {\bibfnamefont {Di}~\bibnamefont
  {Luo}}, \bibinfo {author} {\bibfnamefont {Giuseppe}\ \bibnamefont {Carleo}},
  \bibinfo {author} {\bibfnamefont {Bryan~K.}\ \bibnamefont {Clark}}, \ and\
  \bibinfo {author} {\bibfnamefont {James}\ \bibnamefont {Stokes}},\ }\bibfield
   {title} {\enquote {\bibinfo {title} {Gauge {E}quivariant {N}eural {N}etworks
  for {Q}uantum {L}attice {G}auge {T}heories},}\ }\href {\doibase
  10.1103/physrevlett.127.276402} {\bibfield  {journal} {\bibinfo  {journal}
  {Physical Review Letters}\ }\textbf {\bibinfo {volume} {127}},\ \bibinfo
  {pages} {276402} (\bibinfo {year} {2021})}\BibitemShut {NoStop}%
\bibitem [{\citenamefont {Favoni}\ \emph {et~al.}(2022)\citenamefont {Favoni},
  \citenamefont {Ipp}, \citenamefont {Müller},\ and\ \citenamefont
  {Schuh}}]{Favoni2022}%
  \BibitemOpen
  \bibfield  {author} {\bibinfo {author} {\bibfnamefont {Matteo}\ \bibnamefont
  {Favoni}}, \bibinfo {author} {\bibfnamefont {Andreas}\ \bibnamefont {Ipp}},
  \bibinfo {author} {\bibfnamefont {David~I.}\ \bibnamefont {Müller}}, \ and\
  \bibinfo {author} {\bibfnamefont {Daniel}\ \bibnamefont {Schuh}},\ }\bibfield
   {title} {\enquote {\bibinfo {title} {Lattice {G}auge {E}quivariant
  {C}onvolutional {N}eural {N}etworks},}\ }\href {\doibase
  10.1103/physrevlett.128.032003} {\bibfield  {journal} {\bibinfo  {journal}
  {Physical Review Letters}\ }\textbf {\bibinfo {volume} {128}},\ \bibinfo
  {pages} {032003} (\bibinfo {year} {2022})}\BibitemShut {NoStop}%
\bibitem [{\citenamefont {Apte}\ \emph {et~al.}(2024)\citenamefont {Apte},
  \citenamefont {Córdova}, \citenamefont {Huang},\ and\ \citenamefont
  {Ashmore}}]{Apte2024}%
  \BibitemOpen
  \bibfield  {author} {\bibinfo {author} {\bibfnamefont {Anuj}\ \bibnamefont
  {Apte}}, \bibinfo {author} {\bibfnamefont {Clay}\ \bibnamefont {Córdova}},
  \bibinfo {author} {\bibfnamefont {Tzu-Chen}\ \bibnamefont {Huang}}, \ and\
  \bibinfo {author} {\bibfnamefont {Anthony}\ \bibnamefont {Ashmore}},\
  }\bibfield  {title} {\enquote {\bibinfo {title} {Deep learning lattice gauge
  theories},}\ }\href {\doibase 10.1103/physrevb.110.165133} {\bibfield
  {journal} {\bibinfo  {journal} {Physical Review B}\ }\textbf {\bibinfo
  {volume} {110}},\ \bibinfo {pages} {165133} (\bibinfo {year}
  {2024})}\BibitemShut {NoStop}%
\bibitem [{\citenamefont {Linsel}\ \emph {et~al.}(2024)\citenamefont {Linsel},
  \citenamefont {Bohrdt}, \citenamefont {Homeier}, \citenamefont {Pollet},\
  and\ \citenamefont {Grusdt}}]{Linsel2024}%
  \BibitemOpen
  \bibfield  {author} {\bibinfo {author} {\bibfnamefont {Simon~M.}\
  \bibnamefont {Linsel}}, \bibinfo {author} {\bibfnamefont {Annabelle}\
  \bibnamefont {Bohrdt}}, \bibinfo {author} {\bibfnamefont {Lukas}\
  \bibnamefont {Homeier}}, \bibinfo {author} {\bibfnamefont {Lode}\
  \bibnamefont {Pollet}}, \ and\ \bibinfo {author} {\bibfnamefont {Fabian}\
  \bibnamefont {Grusdt}},\ }\bibfield  {title} {\enquote {\bibinfo {title}
  {Percolation as a confinement order parameter in $\mathbb{Z}_2$ lattice gauge
  theories},}\ }\href {\doibase 10.1103/physrevb.110.l241101} {\bibfield
  {journal} {\bibinfo  {journal} {Physical Review B}\ }\textbf {\bibinfo
  {volume} {110}},\ \bibinfo {pages} {l241101} (\bibinfo {year}
  {2024})}\BibitemShut {NoStop}%
\bibitem [{\citenamefont {Linsel}\ \emph {et~al.}(2025)\citenamefont {Linsel},
  \citenamefont {Pollet},\ and\ \citenamefont {Grusdt}}]{Linsel2025}%
  \BibitemOpen
  \bibfield  {author} {\bibinfo {author} {\bibfnamefont {Simon~M.}\
  \bibnamefont {Linsel}}, \bibinfo {author} {\bibfnamefont {Lode}\ \bibnamefont
  {Pollet}}, \ and\ \bibinfo {author} {\bibfnamefont {Fabian}\ \bibnamefont
  {Grusdt}},\ }\href {\doibase 10.48550/ARXIV.2504.03512} {\enquote {\bibinfo
  {title} {Independent e- and m-anyon confinement in the parallel field toric
  code on non-square lattices},}\ } (\bibinfo {year} {2025}),\ \Eprint
  {http://arxiv.org/abs/2504.03512} {arXiv:2504.03512 [quant-ph]} \BibitemShut
  {NoStop}%
\bibitem [{\citenamefont {Wu}\ \emph {et~al.}(2012)\citenamefont {Wu},
  \citenamefont {Deng},\ and\ \citenamefont {Prokof’ev}}]{Wu2012}%
  \BibitemOpen
  \bibfield  {author} {\bibinfo {author} {\bibfnamefont {Fengcheng}\
  \bibnamefont {Wu}}, \bibinfo {author} {\bibfnamefont {Youjin}\ \bibnamefont
  {Deng}}, \ and\ \bibinfo {author} {\bibfnamefont {Nikolay}\ \bibnamefont
  {Prokof’ev}},\ }\bibfield  {title} {\enquote {\bibinfo {title} {Phase
  diagram of the toric code model in a parallel magnetic field},}\ }\href
  {\doibase 10.1103/physrevb.85.195104} {\bibfield  {journal} {\bibinfo
  {journal} {Physical Review B}\ }\textbf {\bibinfo {volume} {85}},\ \bibinfo
  {pages} {195104} (\bibinfo {year} {2012})}\BibitemShut {NoStop}%
\bibitem [{\citenamefont {Linsel}\ and\ \citenamefont
  {Pollet}(2025)}]{Linsel2025_Paratoric}%
  \BibitemOpen
  \bibfield  {author} {\bibinfo {author} {\bibfnamefont {Simon~M.}\
  \bibnamefont {Linsel}}\ and\ \bibinfo {author} {\bibfnamefont {Lode}\
  \bibnamefont {Pollet}},\ }\href {https://arxiv.org/abs/2510.14781} {\enquote
  {\bibinfo {title} {Paratoric 1.0-beta: Continuous-time quantum {M}onte
  {C}arlo for the toric code in a parallel field},}\ } (\bibinfo {year}
  {2025}),\ \Eprint {http://arxiv.org/abs/2510.14781} {arXiv:2510.14781
  [quant-ph]} \BibitemShut {NoStop}%
\bibitem [{\citenamefont {Gregor}\ \emph {et~al.}(2011)\citenamefont {Gregor},
  \citenamefont {Huse}, \citenamefont {Moessner},\ and\ \citenamefont
  {Sondhi}}]{Gregor2011}%
  \BibitemOpen
  \bibfield  {author} {\bibinfo {author} {\bibfnamefont {K}~\bibnamefont
  {Gregor}}, \bibinfo {author} {\bibfnamefont {David~A}\ \bibnamefont {Huse}},
  \bibinfo {author} {\bibfnamefont {R}~\bibnamefont {Moessner}}, \ and\
  \bibinfo {author} {\bibfnamefont {S~L}\ \bibnamefont {Sondhi}},\ }\bibfield
  {title} {\enquote {\bibinfo {title} {Diagnosing deconfinement and topological
  order},}\ }\href {\doibase 10.1088/1367-2630/13/2/025009} {\bibfield
  {journal} {\bibinfo  {journal} {New Journal of Physics}\ }\textbf {\bibinfo
  {volume} {13}},\ \bibinfo {pages} {025009} (\bibinfo {year}
  {2011})}\BibitemShut {NoStop}%
\bibitem [{\citenamefont {Xu}\ \emph {et~al.}(2025)\citenamefont {Xu},
  \citenamefont {Pollmann},\ and\ \citenamefont {Knap}}]{Xu2025}%
  \BibitemOpen
  \bibfield  {author} {\bibinfo {author} {\bibfnamefont {Wen-Tao}\ \bibnamefont
  {Xu}}, \bibinfo {author} {\bibfnamefont {Frank}\ \bibnamefont {Pollmann}}, \
  and\ \bibinfo {author} {\bibfnamefont {Michael}\ \bibnamefont {Knap}},\
  }\bibfield  {title} {\enquote {\bibinfo {title} {Critical behavior of
  {F}redenhagen-{M}arcu string order parameters at topological phase
  transitions with emergent higher-form symmetries},}\ }\href {\doibase
  10.1038/s41534-025-01030-z} {\bibfield  {journal} {\bibinfo  {journal} {npj
  Quantum Information}\ }\textbf {\bibinfo {volume} {11}} (\bibinfo {year}
  {2025}),\ 10.1038/s41534-025-01030-z}\BibitemShut {NoStop}%
\bibitem [{\citenamefont {Binder}(1981)}]{Binder1981}%
  \BibitemOpen
  \bibfield  {author} {\bibinfo {author} {\bibfnamefont {K.}~\bibnamefont
  {Binder}},\ }\bibfield  {title} {\enquote {\bibinfo {title} {Finite size
  scaling analysis of {I}sing model block distribution functions},}\ }\href
  {\doibase 10.1007/bf01293604} {\bibfield  {journal} {\bibinfo  {journal}
  {Zeitschrift für Physik B Condensed Matter}\ }\textbf {\bibinfo {volume}
  {43}},\ \bibinfo {pages} {119--140} (\bibinfo {year} {1981})}\BibitemShut
  {NoStop}%
\bibitem [{\citenamefont {Hasenbusch}(2008)}]{Hasenbusch2008}%
  \BibitemOpen
  \bibfield  {author} {\bibinfo {author} {\bibfnamefont {Martin}\ \bibnamefont
  {Hasenbusch}},\ }\bibfield  {title} {\enquote {\bibinfo {title} {The {B}inder
  cumulant at the {K}osterlitz–{T}houless transition},}\ }\href {\doibase
  10.1088/1742-5468/2008/08/p08003} {\bibfield  {journal} {\bibinfo  {journal}
  {Journal of Statistical Mechanics: Theory and Experiment}\ }\textbf {\bibinfo
  {volume} {2008}},\ \bibinfo {pages} {P08003} (\bibinfo {year}
  {2008})}\BibitemShut {NoStop}%
\end{thebibliography}

%%%%%%%%%%%%%%%%%%%%%%%%%%%%%%%%%%%%%
% Supplemental material starts
%%%%%%%%%%%%%%%%%%%%%%%%%%%%%%%%%%%%%
\clearpage
\pagebreak
\newpage
\onecolumngrid

\setcounter{equation}{0}
\setcounter{figure}{0}
\setcounter{table}{0}
\setcounter{page}{1}
%\makeatletter

\renewcommand{\theequation}{S\arabic{equation}}
\renewcommand{\thefigure}{S\arabic{figure}}

\begin{center}
\textbf{\large Supplemental Material: Matter-induced plaquette terms in a \texorpdfstring{$\mathbb{Z}_2$}{Z2} lattice gauge theory}
\end{center}

\section{Mapping of the (2+1)D \texorpdfstring{$\mathbb{Z}_2$}{Z2} LGT to the spin model \label{App_SpinModel}}
We are able to integrate out the bosonic matter degrees of freedom, by considering the Gauss's law in the physical sector \cite{Prosko2017}
\begin{equation}
    \hat{G}_{\vecj} \ket{\psi} = \lef -1 \rgh^{\hat{n}_{\vecj}} \prod_{\left \langle \vecj, \textbf{k} \right \rangle \in +_{\vecj}} \tauX_{\left \langle \vecj, \textbf{k} \right \rangle} \ket{\psi} = + \ket{\psi}, \quad  \forall \vecj .
\end{equation}
This directly relates the matter on-site density operator to the configuration of the \Zt electric fields
\begin{equation}
    \hat{n}_{\vecj} = \frac{1}{2} \lef 1 - \prod_{\ijB \in +_{\vecj}} \tauX_{\ijB}  \rgh = \frac{1}{2} \lef 1 - \tauX_{\langle \vecj - \textbf{e}_x, \vecj \rangle} \tauX_{\langle \vecj, \vecj + \textbf{e}_x \rangle} \tauX_{\langle \vecj - \textbf{e}_y, \vecj \rangle} \tauX_{\langle \vecj, \vecj + \textbf{e}_y \rangle} \rgh.
\end{equation}
Using the above relation, we can rewrite the Hamiltonian in Eq.~(1) of the main text as:
\begin{equation}
    \H_s = -\frac{t}{2} \sum_{\ijB } \left [ \tauZ_{\ijB} \lef 1 - \hat{P}_{\textbf{i}} \hat{P}_\textbf{j} \rgh \right ]
    - h \sum_{\ijB } \tauX_{\ijB}
    - J \sum_{\square} \prod_{\ijB \in \square} \tauZ_{\ijB}
    - \frac{\mu}{2} \sum_{\vecj } \prod_{\ijB \in _{\vecj}} \tauX_{\ijB},
    \label{eq_2DLGTwithSpins}
\end{equation}
where we define $\hat{P}_{\textbf{j}} = \prod_{\ijB \in +_\vecj} \tauX_{\ijB}$, and added the plaquette term for completeness.
Since the Hamiltonian in Eq.~(1) of the main text conserves the total parton number $N^{a} = \sum_{\vecj} \hat{n}_{\vecj} = \sum_{\vecj} \ad_{\vecj} \a_{\vecj} $, hopping of a parton can only occur when one of the sites is empty and the other site is occupied.
For example, hopping of a parton from site $\vecj$ to site $\veci$, mediated through the Hamiltonian term $-t\ad_{\veci} \a_{\vecj}$, can only occur when site $\veci$ is empty and site $\vecj$ is occupied by a boson.
This means that we need to consider only the states where $\frac{1}{2} \lef 1 + P_{\veci} \rgh \frac{1}{2} \lef 1 - P_{\vecj} \rgh \ket{\psi} = \ket{\psi}$.
The opposite case is also possible where a parton hops from site $\veci$ to site $\vecj$, which means that we have to consider states where $\frac{1}{2} \lef 1 + P_{\vecj} \rgh \frac{1}{2} \lef 1 - P_{\veci} \rgh \ket{\psi} = \ket{\psi}$.
Combining both cases gives us the projector $\lef 1 - \hat{P}_{\veci} \hat{P}_{\vecj} \rgh$, which only projects to the allowed states where a parton can hop between sites $\veci$ and $\vecj$.
To ensure the actual hopping of the parton between the two neighboring sites, we need to flip the \Zt electric field on the link, which is achieved by the operator $\tauZ_{\ijB}$.

We note that a similar mapping was performed in Ref.~\cite{Borla2022}, where fermionic matter was integrated out using the Gauss's law.
That mapping had to take into account the Jordan-Wigner strings in order to correctly capture the fermionic nature of the matter, and thus contains additional terms.

\section{Numerical simulations \label{App_Numerics}}

\begin{figure}[t]
    \centering
    \epsfig{file=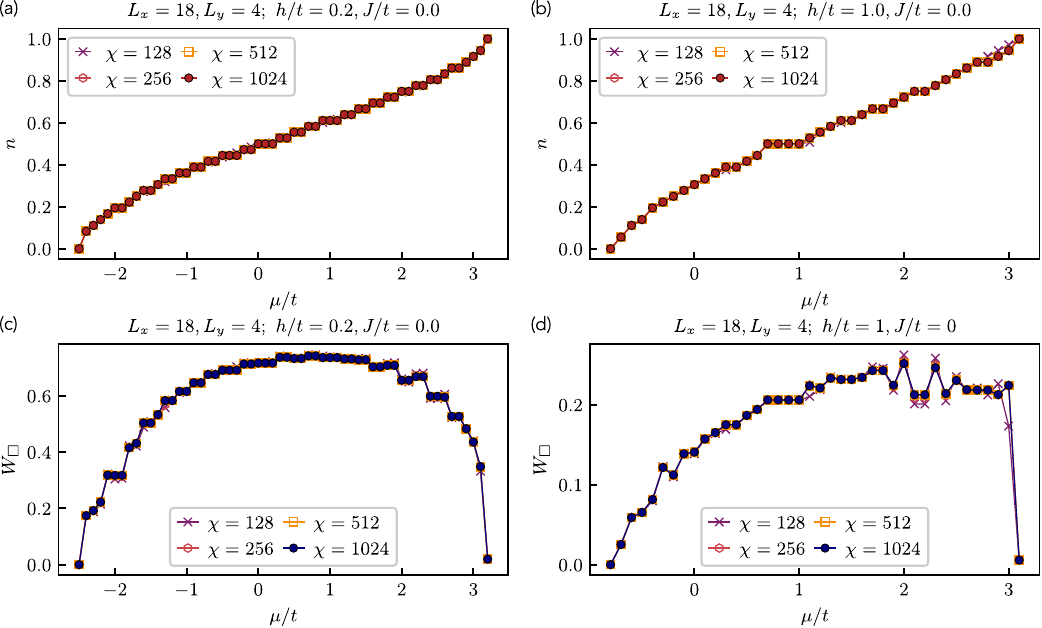}
    \caption{Results of the DMRG calculations of the \Zt LGT Eq.~\eqref{eq_2DLGTwithSpins} for different bond dimension $\chi$. (a) Filling for $h /t = 0.2$ as a function of the chemical potential converges fast already for low bond dimension $\chi \gtrsim 256$ for most value of $\mu$. (b) For larger value of the confining field $h/t = 1$, convergence of the filling $n$, is also fast for relatively small bond dimension $\chi \gtrsim 256$.
    (c) The plaquette values for low field $h / t = 0.2$, also converge nicely, and curves for $\chi  = 512$, and  $\chi  = 1024$ appear to be the same.}
    (d) For higher field $h$, plaquette values converges slower for higher $\mu$, where we still see some slight difference between $\chi = 512$ and $\chi = 1024$.
    \label{App_DMRG_ConvCurves}
\end{figure}

\subsection{DMRG calculation of the ground state \label{App_DMRG}}
The DMRG calculations presented in the main text were performed for the spin Hamiltonian Eq.~\eqref{eq_2DLGTwithSpins} on a cylinder with length $L_x = 18$, and circumference $L_x = 4$.
To be more precise, we considered a square lattice with dimensions $L_x$ along the $x$-direction and $L_y$ along the $y$-direction, and considered open boundary conditions (OBC) in the $x$-direction and periodic boundary conditions (PBC) in the $y$-direction.
We note that the total length of the spin chain used to represent the 2D system corresponds to $L_{\rm MPS} = 2 L_xL_y + L_y$.
DMRG calculations on a cylinder require slightly more care than in the simple one-dimensional case as the NN interactions on the lattice are in fact long-range interactions in the MPS chain, which has to wind through a two-dimensional system.
Due to the unfavorable area laws in the two dimensions, we are severely limited in the system size of one of the spatial directions \cite{Schollwoeck2011} and thus consider $L_x \gg L_y$.
However, as a result we can employ PBC in the shorter direction as the interactions due to the PBC are of the same long-range nature on the MPS lattice, as the interactions that emerge in the bulk of the system.

All of these factors contribute to a generally more difficult convergence compared to a one-dimensional system, and higher bond dimensions $\chi$ are needed.
In order to gain some insight into how well our results converged, we analyze how much our physical observables change by increasing the bond dimension $\chi$.
The main intuition behind this is that the observables converge to the ground state value, with increasing bond dimension as we truncate less singular values and capture more entanglement \cite{Schollwoeck2011}.
Ideally, we thus want to find a large enough $\chi$, which will contain the necessary entanglement and accurately describe our ground state and the physical observables.
We thus need to search for a large enough $\chi$, for which the physical observables do not change significantly by increasing the $\chi$ further.

We perform the DMRG calculations in four stages, where we increase the bond dimension limit in every stage, $\chi = 128, 258, 512, 1024$.
In order to check for convergence, we analyze the physical observables calculated after each stage at different bond dimensions, $\chi$.
In Fig.~\ref{App_DMRG_ConvCurves} we plot the results for the filling $n$ and the plaquette value $W_{\square}$ for different values of the bond dimension limit $\chi$.
We note that these are the same results as in Fig.~2 of the main text where we only show the results obtained after the last stage with the largest bond dimension limit set to $\chi = 1024$.
We see that for the filling $n$, curves at different bond dimensions generally coincide already for $\chi \gtrsim 128$ indicating good convergence.
For higher \Zt electric field value, $h/t =1$, we see that the convergence is slightly worse for the plaquette values for higher chemical potential $\mu$, see Fig.~\ref{App_DMRG_ConvCurves}(d).
There, we see some slight discrepancy even between curves $\chi = 256$ and $\chi = 512$.
However, the curves for $\chi = 512$ and $\chi = 1024$ generally agree well, except for a few data points at high $\mu$, where a slight discrepancy can be seen.

\begin{figure}[t]
    \centering
    \epsfig{file=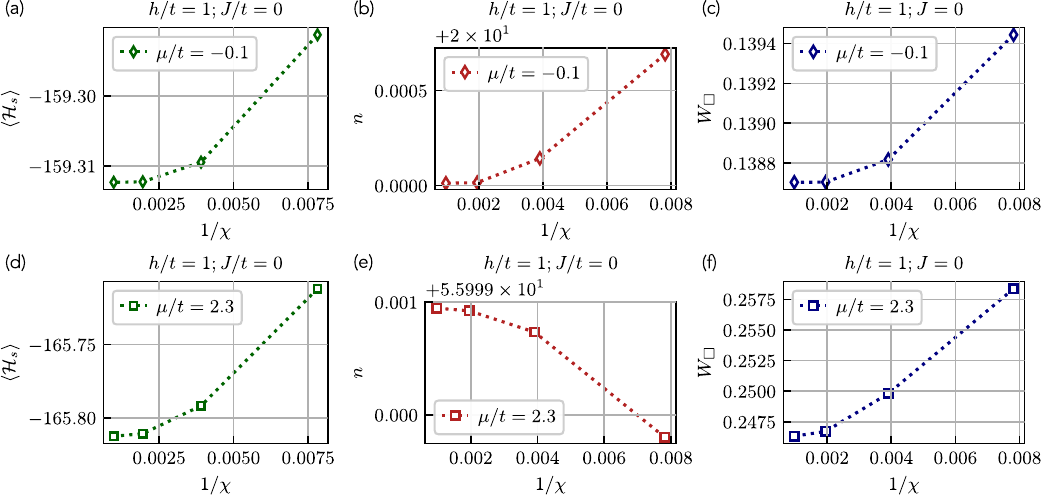}
    \caption{Convergence of the expectation value of the Hamiltonian in Eq.~\eqref{eq_2DLGTwithSpins}, $\langle \H_s \rangle$, lattice filling $n$, and the plaquette term, $W_{\square}$ as a function of inverse bond dimension $1 / \chi$.
    (a)-(c) For the chemical potential $\mu = -0.1t$ we observe good convergence with minimal change for the two largest bond dimensions.
    (d)-(e) We also observe convergence of the observables for the higher chemical potential $\mu$, however the saturation is slightly slower. The electric field term value was $h = t$, and the system size is $L_x = 18$, $L_y =4$.}
    \label{App_DMRG_chi_Dep}
\end{figure}

We analyze this further in Fig.~\ref{App_DMRG_chi_Dep}, where we plot the values of the ground state energy $E$, the filling $n$, and the plaquette value $W_\square$, as a function of the inverse value of the bond dimension limit, $1/ \chi$.
In Figs.~\ref{App_DMRG_chi_Dep}(a)-(c), we plot these quantities for the calculation at $\mu /  t = -0.1$, which is well converge as the values plateau with the increasing value of the bond dimension limit, $\chi$.
We also see a similar, but slightly slower, trend for $\mu /  t = 2.3$, where some discrepancy between the curves could be seen in Fig.~\ref{App_DMRG_ConvCurves}(d).
There, the plateau is not as flat, but the observables clearly start to converge with increasing $\chi$.
Moreover, we believe that the errors are well contained, and we concluded that we would not gain significant advantage by using an even larger bond dimension.
As we are not interested in the exact numerical values, we believe that increasing numerical precision any further would not justify a significant increase of the computational resources and time.
We note that we consider plaquettes $W_\square$ with the lower left vertex of the square at $x = 8$ and $y = 2$.

\subsection{L-CNN Neural quantum state architecture \label{App:Numcalc:NQS}}

For the numerical simulation of $\mathbb{Z}_2$ gauge fields coupled to dynamical matter, we employ a lattice convolutional neural network (L-CNN) model inspired by Luo et al. \cite{Luo2021} and Favoni et al. \cite{Favoni2022, Apte2024}. We also tested several other architectures, such as the vision transformer \cite{Viteritti2023}, which we found to perform worse. In the following, we review the standard NQS ansatz and describe the details of our network architecture.

The general idea of neural quantum states is to represent a variational wave function using machine learning algorithms. To do this, the algorithm uses a configuration $|n,\sigma \rangle$ and predicts its phase $\phi_\lambda(n,\sigma)$ and amplitude $\sqrt{P_\lambda(n,\sigma)}$, with $\lambda$ denoting the dependence on the network parameters. The full wave function is then given by:
\begin{equation}
    |\Psi\rangle = \sum_n \sum_\sigma \psi_\lambda(n,\sigma)|n,\sigma \rangle,
\end{equation}
where $\psi_\lambda(n,\sigma) = \sqrt{P_\lambda(n,\sigma)} e^{i\phi_\lambda(n,\sigma)}$. To optimize the network parameters, we employ stochastic reconfiguration - a linearized imaginary time evolution performed in parameter space.

The $\mathbb{Z}_2$ lattice gauge theory under consideration is characterized by connected and disconnected links. Due to Gauss's law, open strings of connected links can only appear if a particle is located at each end of the string. Inspired by the approaches of Luo et al. \cite{Luo2021} and Favoni et al. \cite{Favoni2022}, we designed a neural network architecture that reflects the physical structure of the system to model the wave function more efficiently.
Specifically, the network transforms the input configuration of strings to capture both local and non-local patterns. The architecture, depicted in Fig.~\ref{App:Fig_NQS}, consists of three key components:

\begin{figure}[t]
    \centering
    \epsfig{file=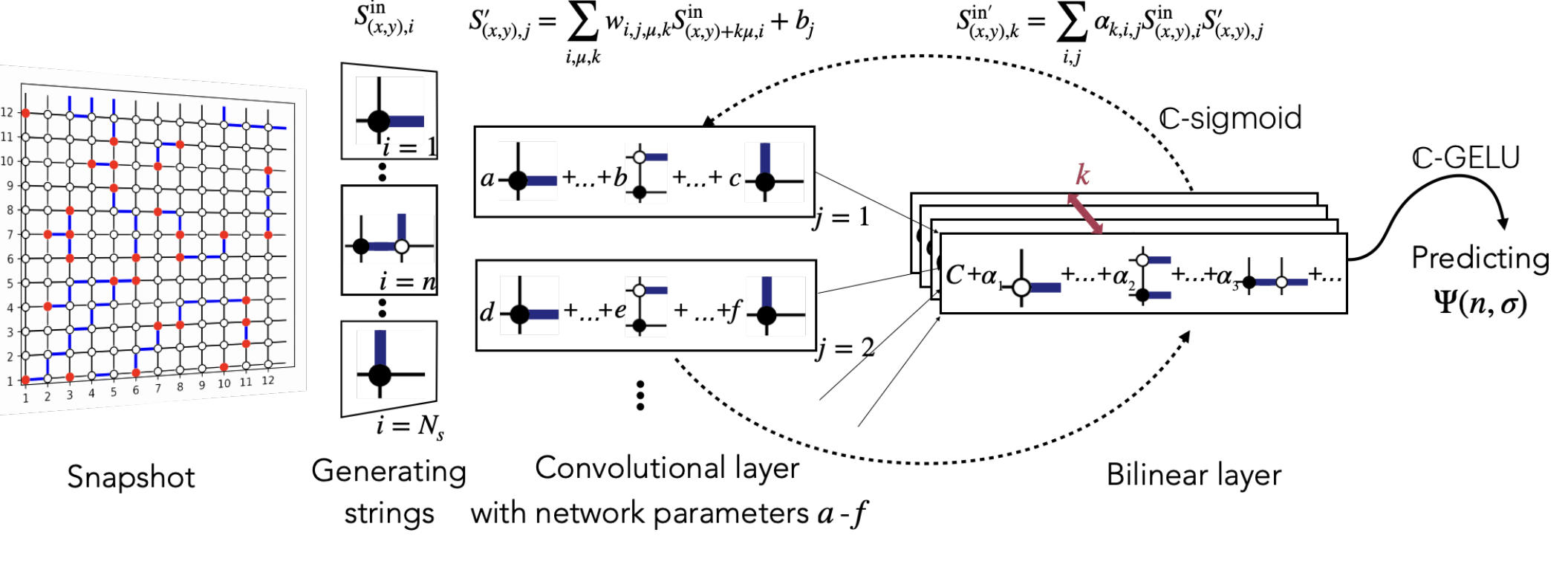, width=0.96\textwidth}
    \caption{Neural quantum state architecture: Before passing the information to the network, each site of a given snapshot is converted into a string of varying length. Thereafter, the network processes the snapshot by using multiple layers of convolutional networks followed by bilinear multiplications. Finally, the wave function coefficient is predicted by applying a linear layer to the output of last bilinear layer.}
    \label{App:Fig_NQS}
\end{figure}

\begin{enumerate}
    \item Before applying any trainable parameters, the input is restructured into a $S^{(N_x \times N_y \times N_s)}$ tensor of input strings:
\begin{equation}
    S^\mathrm{in}_{(x,y),i} = \prod_{j\in l_i} \sigma_{(x,y) + j},
\end{equation}
where $l_i$ denotes the relative path for the $i^\mathrm{th}$ string (with $1 \leq i \leq N_s$), with respect to the lattice site $(x, y)$.

\item Thereafter, we apply a star-shaped convolutional neural network (CNN) filter:
\begin{equation}
    S^{'}_{(x,y),i}  = \sum_{j,\mu,k} w_{i,j,\mu,k} S^\mathrm{in}_{(x,y)+k\cdot \mu,j} + b_i,
\end{equation}
where $i$ and $j$ index the output $N_c$ and input $N_s$ channels, respectively, $\mu\in \{(1,0), (0,1)\}$ specifies the direction, and $-K\leq k \leq K$ defines the position within the convolutional kernel of size $K$.
The resulting string tensor $S^{'(N_x \times N_y \times N_{s'})}$ encodes a weighted sum of various translated string configurations.

\item To modify the effective length and complexity of the strings, a bilinear layer combines them pairwise:
\begin{equation}
        S^{\mathrm{in}'}_{(x,y),i\leq N_s'}  = \sum_{j,k} \alpha_{i,j,k} S^\mathrm{in}_{(x,y),j}S'_{(x,y),k}.
\end{equation}
Here, $w_{i,j,\mu,k}$, $b$, and $\alpha_{i,j,k}$ are trainable network parameters. Following Favoni et al. \cite{Favoni2022}, we combine the convolution and bilinear layers and apply a complex-valued sigmoid activation function to the output. Step 2. and 3. can be repeated several times (with $S^{\mathrm{in}'}_{(x,y),i}$ as the new input with $N_s'$ channels) to increase the effective string length. The final string tensor $S^f$ is then passed through a feedforward network with a complex Gaussian Error Linear Unit ($\mathbb{C}$-GELU) activation function:
\begin{equation}
    \log \psi(n,\sigma) = \mathrm{GELU}\left(\sum_{x,y,i} d_{x,y,i}S^f_{(x,y),i} +b_d\right),
\end{equation}
which predicts logarithm of the wave function coefficient $ \psi(n,\sigma)$.
\end{enumerate}

\begin{center}
\begin{table}[b]
    \centering
    \begin{tabular}{ |c|c|c|c|c|c|} 
    \hline
    $L\times L$ &$6\times 6$ &$8\times 8$ &$10\times 10$ &$12\times 12$ &$20\times 20$ \\
    \hline
    $N^a$ &$8$ &$14$ &$20$ &$30$ &$80$ \\
    \hline
    $n_\mathrm{params}$ &2325 &2409 &2517& 2649 & 3417\\
    \hline
    \end{tabular}
    \caption{Numerical details for the NQS simulations. We fix the density at approximately $20\%$ and tune the electric field $h$. All NQS simulations use about $n_\mathrm{params}\approx\mathcal{O}(10^3)$ variational parameters.}
\label{NQS_Num_details}
\end{table}
\end{center}

\subsubsection{Numerical details} The neural quantum states calculation presented in the main text were performed in on $L \times L$ square lattices with periodic boundary conditions in both directions and range from $L = 6$ with $N^a = 8$ particles to $L = 20$ with $N^a=80$ particles (see Tab. \ref{NQS_Num_details}).
For all simulations, we used the same network architecture with approximately $n_\mathrm{params} = \mathcal{O}(10^3)$ variational parameter. The network produces 
$N_s=6$ different initial strings, followed by ten CNN–bilinear blocks. Each of the ten CNN–bilinear blocks uses the corresponding entries of $N_c= (2,3,4,5,4,3,2,2,2,2)$ and $N'_s= (4,6,8,8,6,6,6,5,5,4,3,3,3,3)$. i.e., block $i$ uses 
$N_c(i)$ channels and produces an output of depth $N'_s(i)$.

\begin{center}
\begin{table}%[]
    \centering
    \begin{tabular}{ |c|c|c|c|c|c|c|c|c|c|c| } 
    \hline
      %Parameters &  &  &  &   \\ 
     %\hline
     $L_x^{\rm OBC} \times L_y^{\rm PBC}$ & $5 \times 4$ & $5 \times 4$ & $5 \times 4$ & $5 \times 4$ & $5 \times 4$  & $5 \times 4$ \\ 
     \hline
     $h/t$ & 0.0 & 0.1 &  0.2 & 0.3 &  0.4 & 0.5  \\ 
     \hline
      $n$ & 8/20 & 8/20 & 8/20 & 8/20 & 8/20& 8/20 \\ 
     %\hline
     % $E_0^{\mu}$ & $-21.144125$ & $-44.99567$ & $-44.639090$ & $-23.185334$ & $-28.544131$ \\ 
     \hline
     $E_0$ & $-19.545$ & $-20.771$ & $-22.756$ &  $-25.189$ & $-27.853$ & $-30.658$ \\ 
     \hline
      $E_0^{\rm NQS}$ & $-19.544 \pm 0.001$  & $-20.755 \pm 0.005$ & $-22.757 \pm 0.003$ &  $-25.185 \pm 0.004$ & $-27.849 \pm 0.005$ & $-30.654 \pm 0.004$\\ 
     \hline
     $W_\square$ & $1.0$ & $0.8895$ & $0.691$ &  $0.554$ & $0.465$ & $0.402$ \\ 
     \hline
      $W_\square^{\rm NQS}$ & $1.0\pm 0.00001$  & $0.895 \pm 0.006$ & $0.696 \pm 0.006$ &  $0.552 \pm 0.005$ & $0.474 \pm 0.008$ & $0.408 \pm 0.006$\\ 
     \hline
    \end{tabular}
    \caption{Comparison of the DMRG and NQS with transformer architecture results. We compare the ground state energy $E_0$ and the plaquette term $W_\square$ for $n = 0.4$ and the \Zt electric field term values $0.0 \leq h \leq 0.5$. The system size is $L_x = 5$, $L_y = 4$. We use open boundary conditions in the $x$-direction and periodic boundary conditions along the $y$-direction.}
    \label{tab_DMRGvsNQS_t1}
\end{table}
\end{center}

\begin{center}
\begin{table}%[]
    \centering
    \begin{tabular}{ |c|c|c|c|c|c|c|c|c|c| } 
    \hline
      %Parameters &  &  &  &   \\ 
     %\hline
     $L_x^{\rm OBC} \times L_y^{\rm PBC}$ & $5 \times 4$ & $5 \times 4$ & $5 \times 4$ & $5 \times 4$ & $5 \times 4$  \\ 
     \hline
     $h/t$ & 0.6 & 0.7 &  0.8 & 0.9 &  1.0   \\ 
     \hline
      $n$ & 8/20 & 8/20 & 8/20 & 8/20 & 8/20\\ 
     \hline
     $E_0$ & $-33.561$ & $-36.539$ & $-39.577$ &  $-42.665$ & $-45.796$ \\ 
     \hline
      $E_0^{\rm NQS}$ & $-33.553  \pm 0.005$  & $-36.52 \pm 0.02$ & $-39.574 \pm 0.004$ &  $-42.663 \pm 0.004$ & $-45.794 \pm 0.005$\\ 
     \hline
     $W_\square$ & $0.354$ & $0.316$ & $0.284$ &  $0.258$ & $0.235$ \\ 
     \hline
      $W_\square^{\rm NQS}$ & $0.360\pm 0.008$  & $0.318 \pm 0.006$ & $0.291 \pm 0.007$ &  $0.255 \pm 0.006$ & $0.234 \pm 0.006$ \\ 
     \hline
    \end{tabular}
    \caption{Comparison of the DMRG and NQS with transformer architecture results. Same observables and parameters as in Table~\ref{tab_DMRGvsNQS_t1} for larger \Zt electric field term values $0.6 \leq h \leq 1.0$.}
    \label{tab_DMRGvsNQS_t2}
\end{table}
\end{center}

\subsubsection{Calculating error bars} To quantify the statistical uncertainty in our neural quantum state calculations, we use the standard error of the mean as the default error estimate for observables such as the energy, the plaquette term and the pair-pair correlations. For the Percolation based order parameters and the Fredenhagen-Marcu, we used a bootstrap error estimation.

\begin{center}
\begin{table}%[]
    \centering
    \begin{tabular}{ |c|c|c|c|c|c|c| } 
    \hline
 $h/t$& $0.1$&$0.2$& $0.3$ & $0.4$ & $0.5$ & $0.6$\\
  \hline
 L-CNN &  $-28.386 \pm 0.002$  & $-32.578 \pm 0.004 $ & $-37.541 \pm 0.003$  & $-42.794 \pm 0.005$ &$-48.221 \pm 0.004$ & $-53.765 \pm 0.004$\\
  \hline
 ViT & $-28.378 \pm 0.005$ & $-32.569 \pm 0.008$ &    $-37.538 \pm 0.005$ & $-42.791 \pm 0.004$ & $-48.211 \pm 0.005$ & $-53.752 \pm 0.006$ \\
 \hline
    \end{tabular}
    \caption{Comparison of the NQS energy with L-CNN and ViT architecture for a $6\times 6$ square lattice with $N=8$ particles and periodic boundary conditions in both directions.}
    \label{tab_LCNN_ViT}
\end{table}
\end{center}

\subsection{Benchmarks of the DMRG and the NQS \label{App_NumBench}}
Here we present the benchmark calculations where we use the DMRG results to verify the NQS calculations.
We directly compare the ground state energy $E_{0}$ and the plaquette term value $W_{\square}$, for different parameter values.
The values are presented in Table~\ref{tab_DMRGvsNQS_t1} and Table~\ref{tab_DMRGvsNQS_t2}.
The system size is $L_x = 5$, $L_y = 4$, with periodic boundary conditions in the $y$-direction, and the filling is fixed to $n = 0.4$.
We see that the values of the energy, $E_0$ and the plaquette term $W_{\square}$, lie within the error margins defined by the standard deviation of the mean.

For the DMRG results, we note that we subtract the chemical potential term $\propto \mu \sum_{\vecj} \hat{n}_{\vecj}$, from the expectation value of the Hamiltonian in Eq.~\eqref{eq_2DLGTwithSpins}.
We do this to obtain the ground state energy that can be directly compared with the NQS results where the filling was directly fixed to the desired value.

We note that our specific NQS architecture is designed for systems with periodic boundary conditions, since it uses strings that cross boundaries. Therefore, our architecture naturally struggles with open boundary conditions, which are necessary for a comparison with DMRG. To overcome this hurdle, we use a standard vision transformer NQS \cite{Viteritti2023} for the benchmark, which we found to perform worse than the L-CNN network for systems with periodic boundary conditions; see the comparison in Table~\ref{tab_LCNN_ViT}.

In contrast, DMRG calculations with PBC in both spatial dimensions proved to be very difficult, and could only be converged for very small system sizes.
Thus, both methods are complementary, with DMRG being the method of choice on mixed-dimensional systems where $L_x \gg L_y$, with OBC in the longer dimension and PBC in the shorter dimension, for generic filling and parameter values.
However, the L-CNN NQS is the method to use in large two-dimensional systems with PBC in both spatial dimension, but encounter difficulties in the OBC.
We note that in the case of $J < 0$, the ground state can exhibit a complicated sign structure, which can, in general, make the NQS convergence more difficult.

\subsection{DMRG results for the Fredenhagen-Marcu order parameter}\label{App_FHforDMRG}
\begin{figure}[t]
    \centering
    \epsfig{file=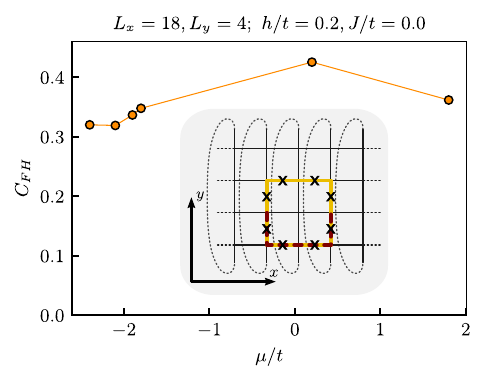}
    \caption{Fredenhagen-Marcu order parameter Eq.~\eqref{eq_FHorderParameter} as a function of the chemical potential $\mu$ for $h / t = 0.2$. We used DMRG on a cylinder of length $L_x = 18$, and circumference $L_y = 4$. The sketch depicts the full loop (yellow loop, $\mathcal{C}$) and the half loop (red string, $\mathcal{C}_{1/2}$) of the operators $\prod  \tauZ$, which form the Fredenhagen-Marcu order parameter.}
    \label{Fig_FH_DMRG_h02}
\end{figure}

To probe confinement of partons on the cylinder of length $L_x = 18$ and circumference $L_y=4$ for $h / t = 0.2$ in our DMRG calculations,  we compute the Fredenhagen-Marcu order parameter defined as \cite{Fredenhagen1983, Fredenhagen1986, Marcu1986, Fredenhagen1988, Gregor2011, Xu2025}
\begin{equation}
    C_{FH} = \frac{ \left \langle  \prod_{\ijB \in \mathcal{C}_{1/2}} \tauZ_{\ijB} \right \rangle }{\sqrt{\left \langle  \prod_{\ijB \in \mathcal{C}} \tauZ_{\ijB} \right \rangle}}.
    \label{eq_FHorderParameter}
\end{equation}
Here, $\mathcal{C}$ is the square path with dimension $\ell \times \ell$, and $\mathcal{C}_{1/2}$ is the half of such loop; see the sketch in Fig.~\ref{Fig_FH_DMRG_h02}.
We note that a simple plaquette term in the main text corresponds to the loop of dimension $\ell \times \ell = 1 \times 1$.
For $h =0.2$ and the system size $L_x = 18 \gg L_y =4$, we find that the DMRG calculations yield a finite Fredenhagen-Marcu order parameter, which points to partons already confining into mesons; see Fig.~\ref{Fig_FH_DMRG_h02}.
However, we note that the cylinder geometry complicates the analysis and the computation of the Fredenhagen-Marcu order parameter, as it limits the maximum length of the loop to $\ell = 2$.
The lower left vertex of the loop $\mathcal{C}$ presented in Fig.~\ref{Fig_FH_DMRG_h02} is at $x= 8$ and $y = 1$.

\subsection{NQS results for the Binder cumulant and the Fredenhagen-Marcu order parameter}
\label{App_NQS_Binder_FM}

\begin{figure}[t]
    \centering
    \epsfig{file=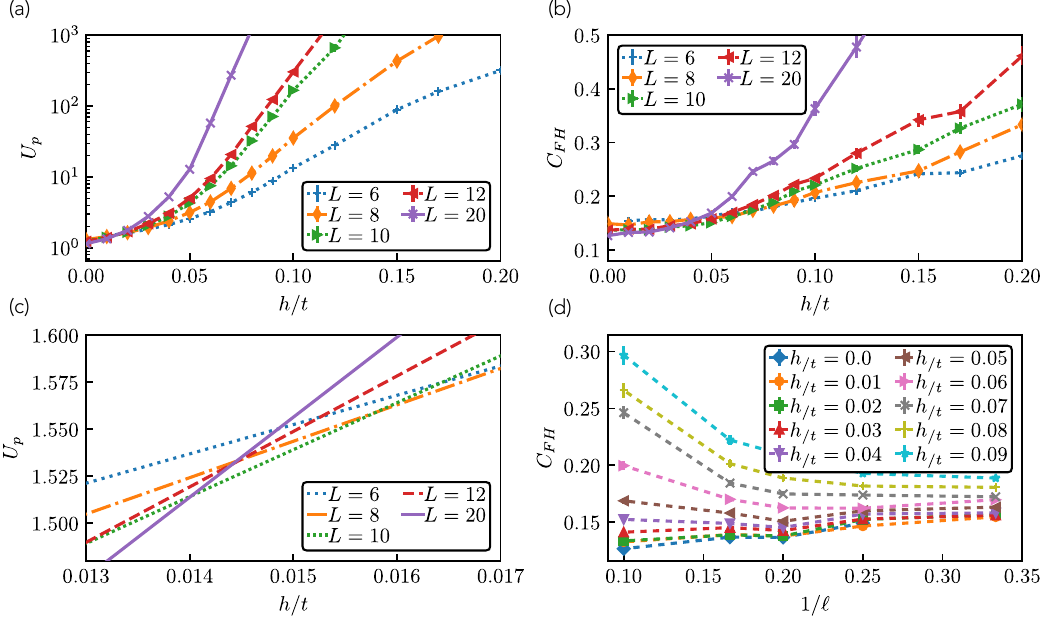}
    \caption{Binder cumulant $U_p$ and Fredenhagen-Marcu order parameter $C_{FH}$ for different system sizes $L\times L$ and electric field strengths $h/t$. Panels (a) and (c) show $U_p$, with (c) providing a close up of the crossing region near $h/t\approx 0.015$. Although the crossing points are slightly scattered, there is no trend towards smaller $h/t$ values with increasing system size. Panel (b,~d) display $C_{FH}$ as a function of (b) the electric field $h/t$ and (d) the inverse width $\frac{1}{\ell} = \frac{2}{L}$ of the Wilson loop. The response of $C_{FH}$ in (b) and the change of the scaling in $1/\ell$ in (d) around $0.02 \lesssim h/t \lesssim 0.04 $ indicate a possible phase transition in the same region suggested by $U_p$, although larger system sizes are required for a definite conclusion.}
    \label{App_Fig_NQS_Binder_FM}
\end{figure}

In the main text, we observed evidence of a phase transition at $h/t \approx 0.015$ by evaluating the percolation and the Binder cumulant of the percolation strength $U_p = \frac{\langle P^4(h,L)\rangle}{\langle P^2(h,L)\rangle^2}$, where $P(h,L)$ is the size of the percolating cluster \cite{Linsel2024}.
Here, we show more detailed results of the Binder cumulant in Fig.~\ref{App_Fig_NQS_Binder_FM}(a) and Fig.~\ref{App_Fig_NQS_Binder_FM}(c)
In addition, we discuss evidence of a phase transition in the Fredenhagen-Marcu order parameter $C_{FH}$, defined in Eq.~\ref{eq_FHorderParameter} with a loop size of $\ell\times \ell = \frac{L}{2} \times \frac{L}{2}$, in Fig.~\ref{App_Fig_NQS_Binder_FM}(b) and Fig.~\ref{App_Fig_NQS_Binder_FM}(d).

In Fig.~\ref{App_Fig_NQS_Binder_FM}(a) we show the full $h/t$ range of the exponentially scaling Binder cumulant $U_p$ and its crossing point at $h/t \approx 0.015$. Fig.~\ref{App_Fig_NQS_Binder_FM}(c) shows a close up of the transition point.
The crossing points for all system sizes are slightly scattered within the range $0.013 \lesssim h/t \lesssim 0.017$ and $1.49 \lesssim U_p \lesssim 1.58$. 
However, no clear trend towards smaller $h/t$ values with increasing system size can be identified.
Moreover, the spread of the Binder cumulant curves becomes minimal at $h/t \approx 0.0148$, where a single well-defined crossing is only marginally missed, potentially due to numerical inaccuracies, convergence issues, small deviations in the densities ($ \frac{N^a}{L\times L} \approx 20\%$), and finite size effects. Nevertheless, these results provide evidence that a transition at finite $h/t$ values persists in the thermodynamic limit. 

The Fredenhagen-Marcu order parameter $C_{FM}$ plotted as a function of the electric field strength $h/t$ is shown in Fig.~\ref{App_Fig_NQS_Binder_FM}(b).
Here, we observe a significant increase of $C_{FM}$ after the predicted transition point $h/t \approx 0.015$, which, in principle, indicates a confined-deconfined transition.
However, determining the correct onset would require much larger system sizes.
Note, that the $C_{FM}(h/t\rightarrow 0)$ is finite for all simulated systems, which is not expected in the thermodynamic limit $L \rightarrow\infty$.
In Fig.~\ref{App_Fig_NQS_Binder_FM}(d), we plot $C_{FM}^{h/t}$ as a function of the inverse width $1/\ell = 2/L$ of the Wilson loop.
For $h/t > 0.04$, we observe a monotonic decline of $C_{FM}$ as a function of $1/\ell$.
This changes around $0.02 \lesssim h/t \lesssim 0.04 $ to a monotonic increase of  $C_{FM}(1/\ell)$.
The change of the scaling behavior indicates a phase transition in the region $h/t \lesssim 0.05$.
However, to make definite claims, much larger system sizes are required.

\section{Pair-pair correlations \label{App_PairPair}}
In order to investigate the behavior of the mesons across the deconfinement transition, we compute the pair-pair correlations, which we define as
\begin{equation}
    \left \langle \hat{\Delta}^{\dagger}_{\veci} \hat{\Delta}_{\vecj} \right \rangle = \left \langle 
    \lef \ad_{\veci} \tauZ_{\langle \veci, \veci + a \hat{e}_x \rangle} \ad_{\veci + a \hat{e}_x} \rgh
    \lef \a_{\vecj} \tauZ_{\langle \vecj, \vecj + a \hat{e}_x \rangle} \a_{\vecj} \rgh
    \right \rangle ,
    \label{Eq_PairPair}
\end{equation}
where $\hat{e}_x$ is the unit vector along the $x$-direction and $a$ is the lattice spacing.

\begin{figure}%[t]
    \centering
    \epsfig{file=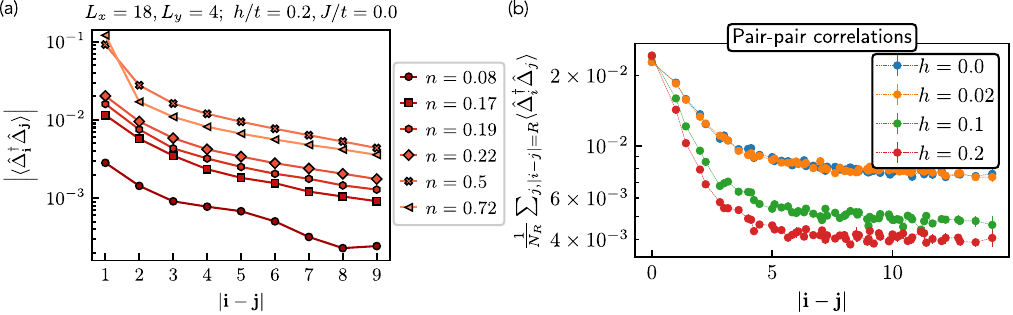}
    \caption{Pair-pair correlations Eq.~\eqref{Eq_PairPair}.
    (a) DMRG results for different filling $n$ in a system size of $L_x = 18$, $L_y = 4$, moving along the $x$-direction for a fixed $i_y = j_y = 3$, i.e. $\veci - \vecj = (i_{x} - j_x, 0)$.
    (b) Pair-pair correlation for the $L_x \times L_y = 20 \times 20$, $n^a = 0.2$ system shown in Fig.~3 of the main text. Here, we average over all $j$'s with the same distance to site $\{i_x=10,i_y = 10\}$.
    }
    \label{Fig_AppPairPair}
\end{figure}

\subsection{DMRG results for the pair-pair correlations}
The DMRG is limited to a cylinder with circumference $L_y = 4$, and the results exhibit a fast decay on the short length scale, which becomes a much slower power-law decay at longer distances; see Fig.~\ref{Fig_AppPairPair}(a).
Here, $\veci - \vecj =  (i_x - j_x, 0)$, and $i_y = j_y=3$.
This is the expected behavior, as the system is not fully two dimensional ($L_x = 18 \gg L_y =4$) and retains some of the one-dimensional features.
We observe the same features across different fillings $n$, but with faster initial decays for higher fillings.

\subsection{NQS results for the pair-pair correlations}
We also compute the pair-pair correlations in the larger two-dimensional system using the NQS.
Fig.~\ref{Fig_AppPairPair}(b) shows NQS results for the pair-pair correlations on a larger, fully two-dimensional system with $L_x = L_y = 20$ and periodic boundary conditions in both directions.
Here, we use a fixed filling $n = 0.2$ and vary the electric field strengths $h$. The NQS results show the saturation of the correlator to a finite value at longer distances $R = | \veci - \vecj |$.
In these calculations, we fixed $\veci = {i_x = 10, i_y = 10}$ and averaged over all sites $\vecj$ with the same distance $R$.

Our results show signatures of the condensation of the \Zt neutral mesons.
However, larger system sizes will be needed for a conclusive claim.

\end{document}